\documentclass[prd,showpacs,superscriptaddress]{revtex4}
\usepackage{epsfig}
\usepackage{mathrsfs}
\usepackage{amsmath}
\usepackage{amssymb}
\usepackage{color}
\usepackage{graphicx}% Include figure files
\usepackage{dcolumn}% Align table columns on decimal point
\usepackage{bm}% bold math

\begin{document}

 \title{Multiple Parton Interactions in
Hadron Collisions and Diffraction}

\author{Paolo Lipari}
\email{paolo.lipari@roma1.infn.it}
\affiliation{INFN sezione 
Roma ``La Sapienza''} 

\author{Maurizio Lusignoli}
\email{maurizio.lusignoli@roma1.infn.it}
\affiliation{Dipartimento di Fisica and sezione INFN,
Universit\`a di Roma ``La Sapienza''}

%\date{February 27, 2000}          

\begin{abstract}
Hadrons are composite objects  made of quarks and  gluons, and during
a collision one can have several elementary interactions  between
the constituents. 
These elementary interactions, 
using an appropriate  theoretical  framework,
can be related to the total and elastic cross sections.
At high  c.m. energy  it also  becomes possible
to identify  experimentally a  high  $p_\perp$  subset of 
the parton interactions and to study their  multiplicity  distribution.
Predictions of the  multiple  interactions  rates 
are  difficult  because  in principle one  needs to have  
a knowledge  of  the correlated  Parton Distribution Functions
that  describe  the probability to  find  simultaneously
different  partons  in different elements of  phase space.
In this work  we  address this  question and  suggest
a  method to describe effectively  the fluctuations
in the instantaneous  configuration of a colliding hadron.
This problem is intimately  related  to  the origin
of the inelastic  diffractive processes.
We present a new method
to include  the  diffractive cross section in an eikonal  formalism
that is  equivalent to a  multi--channel eikonal.
We compare with data  and present an  extrapolation to higher energy.
\end{abstract}


\maketitle

\section{Introduction}
\label{sec:introduction}
The evolution with center of mass energy of the 
total and elastic cross sections in hadron--hadron  collisions,
and the properties of multi--particle production in these interactions
remain an important  open problem in  particle physics.
This  problem is  clearly  of significant intrinsic interest,
but it has  also  important phenomenological implications:
on one hand the  estimate of the properties of  hadronic  interactions
is  obviously important at  LHC to model  the  background 
in the search for   Higgs particles and  possible  forms  of
``new physics'', on the other   hand the 
detailed   properties of  hadronic
interactions properties is  important in 
cosmic ray  studies.
The highest energy  cosmic  ray particles can only be detected  indirectly
 observing the extensive air  showers   that they produce  in the
Earth's atmosphere. The  development of  air showers 
is   determined   by the hadronic  cross sections
and the properties of particle production in hadronic  interactions,
 and therefore    the interpretation of  the available
(and future)  data  depends also  on
theoretical assumptions about   hadronic  interactions.

Hadrons are composite objects containing  quarks  and  gluons, and it
is  natural to  relate the  cross sections for 
hadron  collisions  to the more elementary
interactions  between their  parton  constituents.
The precise method to do this  remains however  an  unsolved problem.

A possible  approach  is  provided by the so  called
``minijet'' models \cite{Durand:1988cr}, where the  total and
elastic cross sections  are obtained using an eikonal formalism,
in  terms  of the quantity  $\langle  n(b,s) \rangle$  that  has  the physical
meaning  of the average  number of elementary  interactions
at  impact parameter $b$ and  c.m. energy $\sqrt{s}$
(for earlier  work on the importance
of minjets see  \cite{Gaisser:1984pg,Pancheri:1985ix}).
An attractive  feature of this  approach is 
that it  allows to compute the
distribution of the number of  elementary  interactions 
that  happen in a single hadron collision.
This  distribution  can then be used, with the inclusion
of a   few additional   assumptions,
in Montecarlo  codes to predict   properties of 
particle  production  that can be  tested  experimentally.
For example it is simple to see that events 
with a large  number of  parton scatterings  must have a 
larger multiplicity and a more complex structure.

The original version of the minijet  approach
and several subsequent ones  did  not  include
in a consistent  way  in their formalism the  inelastic  diffractive  processes.
Following  an approach introduced long ago  by Good  and Walker \cite{Good:1960ba},
several authors 
\cite{Kaidalov:1979jz,Fletcher:1992sy,Roesler:1993yr,Ryskin:2007qx,Gotsman:2008tr}
have  indicated  possible
methods to include  diffraction using a multi-channel  eikonal  formalism.
In this work  we rediscuss  this problem, and
suggest  an alternative  method 
to include   the diffractive  cross section  in the eikonal  formalism.
This method is mathematically equivalent to  the multi--channel
eikonal method  but  offers  additional physical  insight.

The fundamental  physical  idea  to  explain the existence of
inelastic  diffraction introduced 
by Good  and Walker \cite{Good:1960ba} is to assume 
that inelastic  diffraction emerges  because an interacting  hadron
can be seen  as a superposition  of different  states  that  undergo
unequal absorptions.  It is natural, 
as  originally proposed by Miettinen  and Pumplin \cite{Miettinen:1978jb}, 
to identify   these  ``transmission eigenstates''  as  
different  ``configurations''  of the parton  constituents inside  a
hadron.  In this  theoretical  framework
the  estimate  of inelastic  diffraction requires  some
understanding  of the ensemble of  such parton configurations.
This appears as  a daunting task.    

A  possible approach  is to  make the  dramatic  
approximation  of   reducing the space of  parton
configurations  to a finite dimensional space
(in  fact  spanned  by as  few as  two or three  base vectors)
and to construct explicitly  a  matrix transition operator.
An explicit example  (originally constructed in  \cite{Ahn:2009wx})
of  this  approach is also discussed  in this work.
 
The alternative method  we  propose here is to construct a  mapping
from the space of the hadron configurations to the  real  positive 
numbers, so that the  number of  elementary   interactions
for  the configuration ${\mathbb{C}}$ is  $n({\mathbb{C}}) = \langle n\rangle  \, \alpha ({\mathbb{C}})$
(with $\langle n \rangle$ the  average over all configurations).
The probability distribution of $\alpha$  
(together with a model for $\langle n\rangle$) is  then  sufficient
to compute   the total cross section and its different 
components (elastic, absorption and  diffraction). 

The consistent inclusion of  diffraction  in the theory 
is  very important, because it changes  dramatically the relation 
between the  inclusive parton cross  sections
and the directly observable hadron cross sections.

This work is  organized as  follows.
In the next section   we start discussing   the problem
of  multiple interactions  in a narrower  but  better  defined  sense, that 
is limiting our  considerations to  hard  high $p_\perp$  interactions
that on one hand can  be easily identified  experimentally,  and on the other
hand have an inclusive  cross  section that is  calculable in perturbative QCD
using  the standard Parton Distribution Functions (PDF's).
We  will show  that the calculation of the  multiplicity 
distribution of these hard interactions   requires  additional   theoretical
constructs.

In section~\ref{sec:total}   after  rewiewing some general formalism
about total and  elastic cross sections  we  discuss a
standard, single  channel,   version  of the eikonal  formalism,
that has  been  used  in  the  original minijet model  of  Durand and Pi
and in many other  works.

In section~\ref{sec:diff}, after a  brief   introduction to
inelastic  diffraction we  recall   the  basic  ideas  of the 
Good and Walker \cite{Good:1960ba} method, and  then
we  discuss  our   implementation of this multi--component ansatz
in terms  of   the function  $p(\alpha)$  (where $\alpha$  is  a  real  positive 
variable). We call  the  function  $p(\alpha)$  the 
``effective configuration probability  distribution''.

In section~\ref{sec:explicit}  we present a simple form 
for the function $p(\alpha)$ that  depends on  a single  parameter,
and use this  form, together with a  (2 parameters) parametrization 
of the  function $\langle  n(b,s) \rangle$   to 
describe  the  available data 
on $pp$ and $\overline{p}p$  scattering at collider energies.

In section~\ref{sec:energy} we  discuss   the  energy dependence 
of the parameters  of  our  model  and discuss  extrapolations to 
higher  energy (LHC and ultra high energy cosmic  rays).
Section~\ref{sec:concl}  offers some final  considerations.

\section{Multiple  Interactions}
\label{sec:multiple}
The problem of multiple  interactions  in 
a hadron--hadron collisions is  usually  discussed in the context 
of a calculation of the total  (or inelastic)   cross section
and  referring to the {\em  total} number of   elementary interactions
in  a  collision.
Such a general  discussion has serious  difficulties. 
Theoretically the concept of the ``total''   number of   elementary
interactions in a collision  is  not  really well defined.  
For example, one   usually   divides  the elementary
interactions  into  two  classes:   ``soft'' and ``hard''
choosing a rather  arbitrary cutoff in $p_\perp$,  however
``soft''   interactions   cannot  be considered as truly  elementary,
since  they are  effective processes, as  for instance a 
``pomeron exchange'' that, after  decades of efforts,  
remains  a somewhat  elusive  concept;
and the  theoretical ``counting'' of the  number of
interactions in a collision  has  significant  ambiguities.
On the other hand, experimentally  it is  essentially  impossible to
measure the  number of  soft interactions in one collision, and
one can at best obtain  only some  partial and indirect   information
from  the study  of particle   multiplicities.

To avoid  these  difficulties, in this section we  will discuss
a more  limited  but much better  defined  problem,  namely the 
production of   parton--parton 
scatterings    with a  transverse   momentum  
larger  than  a   chosen  threshold. 
If   the  threshold $p_\perp^{\rm min}$, is  sufficiently large 
(above a few  Gev) these  scatterings are, at least in principle,
experimentally identifiable as pairs  of back to  back jets.
Moreover,  given the colliding  hadron PDF's, 
the inclusive differential cross  section  for  the production of 
 pairs of jets  can be  estimated in perturbation theory from  the
well known  expression:
\begin{equation}
\left . \frac{d^3 \sigma}{dp_\perp \, dx_1 \, dx_2}
\right |_{\rm jet~pair} (p_\perp, x_1 , x_2; \sqrt{s})
= 
 \sum_{j, k, j^\prime, k^\prime} ~
~f^{h_1}_j (x_1, \mu^2) 
~
f^{h_2}_k (x_2, \mu^2) 
~\frac{d \hat{\sigma}_{jk\to j^\prime k^\prime}}{dp_\perp} (p_\perp, \hat{s})
\label{eq:jet1}
\end{equation}
In this  expression $f_j^{h_1}(x, \mu^2)$ [$f_k^{h_2}(x, \mu^2) ]$  
is the Parton Distribution Function
(PDF)  for parton  of type  $j$ ($k$)  in   the  hadron $h_1$ ($h_2$)
ar the scale  $\mu^2 \sim  p_\perp^2$,
and $d\hat{\sigma}_{jk \to j^\prime k^\prime}/dp_\perp (p_\perp, \hat{s})$ 
is the  differential  cross  section for  the  parton--parton scattering
of type $j+k \to  j^\prime + k^\prime$ at  the   squared c.m.  energy for the
parton--parton interaction $\hat{s} = s\, x_1 \, x_2$.
The  fractional  momenta $x_{1,2}$ are connected to the 
rapidities $y_{1,2}$  of the observed jets  
by  the relation:
\begin{equation}
x_{1,2} = \frac{p_\perp}{\sqrt{s}} ~\exp \left [ \pm \frac{ (y_1 + y_2)} {2}  \right ]
\end{equation}

Integrating equation  (\ref{eq:jet1})  over the  
phase space region  $p_\perp > p_\perp^{\rm min}$
and  all allowed jet rapidities   one obtains the  inclusive  jet 
 cross section:
\begin{equation}
 \sigma_{\rm jet} (p_\perp^{\rm min}, \sqrt{s}) 
= \int_{p_\perp^{\rm min}}^{\sqrt{s}/2} dp_\perp 
~\int_{4 \, p_\perp^2/s}^1 dx_1
~\int_{4 \, p_\perp^2/(s \, x_1)}^1 dx_2
~\left \{
 \sum_{j, k, j^\prime, k^\prime} ~
~f^{h_1}_j (x_1, \mu^2) 
~
f^{h_2}_k (x_2, \mu^2) 
~\frac{d \hat{\sigma}_{jk\to j^\prime k^\prime}}{dp_\perp} (p_\perp, \hat{s})
\right \}
\label{eq:jet2a}
\end{equation}
More  in general, one could  consider  the   production of jet  pairs
in a more limited    region of  phase space  selecting for  example  only
jets in a certain range of  rapidities, with appropriate  choice
of the limits in the  integration  over phase space. 
The  discussion  below   remains    valid
also in this  case,  and  in the  following we will denote with
$\sigma_{\rm jet}$ the inclusive  production of  jets in
a fixed kinematical region  determined  by $p_\perp^{\rm min}$ and
appropriate  cuts in the jet  rapidities, leaving  the  dependence
on  these  kinematical  cuts  implicit.

The  quantity $\sigma_{\rm jet} (s)$ 
is  a  cross section for  parton--parton   interactions, and therefore 
must be interpreted in  the appropriate way.
Its  physical  meaning is to give the  
{\em inclusive}  cross section for  
the production of  jet--pairs  in the chosen kinematical region.
This  means that when a detector  
collects   the  integrated  luminosity $L_{\rm int}$, 
the    expected  number of  jet pairs  
 is $L_{\rm int} ~\sigma_{\rm jet}$.
Since the total  number of  inelastic   scattering  events is
 $L_{\rm int} ~\sigma_{\rm inel}$,  the ratio
$\sigma_{\rm jet}/\sigma_{\rm inel}$  is   the  average number 
of  jet pairs produced in one inelastic  interaction.
In principle it is possible, and in fact it will happen  for
$p_\perp^{\rm min}$ sufficiently small and/or   $\sqrt{s}$ sufficiently large,
that  $\sigma_{\rm  jet}$ exceeds  $\sigma_{\rm inel}$.
This   simply implies that some 
events  must contain  more than one  parton--parton  interaction.

In general,  an inelastic event can  have 0,1,2  or more
hard interactions. A  natural  problem   is 
the estimate  of the  relative frequencies of 
events   that  have  jet   multiplicity  $k$.
The  probability $p_k$ that an  inelastic  event  contains 
exactly $k$~pairs of  jets  can be expressed as the  ratio:
\begin{equation}
 p_k  = \frac{\sigma_k^{\rm jet}}{\sigma_{\rm inel}}
\end{equation}
%In the last equation and in the following   we leave implicit the 
%dependence on the c.m. energy $\sqrt{s}$ and
%the  threshold   $p_\perp^{\rm surv}$. 
The  partial  cross  sections 
$\sigma_k^{\rm jet}$ must satisfy the sum rules:
\begin{equation}
\sum_k  \sigma_k^{\rm jet} =  \sigma_{\rm inel}
\label{eq:sumj1}
\end{equation}
\begin{equation}
\sum_k  k~\sigma_k^{\rm jet} =  \sigma_{\rm jet}
\label{eq:sumj2}
\end{equation}
and therefore 
\begin{equation}
\langle  k\rangle \equiv 
\sum_k  k~p_k^{\rm jet} = \frac{\sigma_{\rm jet}} {\sigma_{\rm inel}}
\label{eq:kmed}
\end{equation} 

It is important to stress that the  set of partial  cross sections 
$\sigma_k^{\rm jet}$,  or equivalently the  probabilities
$ p_k$  are 
observable  quantities.
For  large $p_\perp^{\rm min}$ the identification of the hard
interactions  is  experimentaly straightforward, 
however  the inclusive jet cross section 
$\sigma_{\rm jet}$
is  much  smaller  than $\sigma_{\rm inel}$ and   the
jet  multiplicity   distribution becomes   ``trivial'' 
and  only $p_0$  and  $p_1$   are  different from zero.
For   $p_\perp^{\rm min}$     sufficiently small 
the  jet multiplicity   distribution
become  broader,
and the  probability to find more
than one  hard scattering in a single event becomes
significant; however  at the same  time the 
experimental identification  of the hard scatterings  becomes  more
difficult.
At LHC  it should  however be  possible
to  identify a   value  $p_\perp^{\rm min}$  
sufficiently small to result in an interesting
multiplicity  distribution  of hard interactions, and sufficiently 
large to  allow the measurement of  such a distribution.

Such an experimental  study   should be compared with a  theoretical
prediction. The   calculation   of the  multiplicity 
distribution of  the hard interactions
is  in fact a  very difficult, unsolved  problem,
that,  as we will  discuss in the following,  
requires the introduction of new ideas, beyond  the use of the standard PDF's, 
that are only  sufficient for the calculation
of the  inclusive jet cross section.
One may note  that  the obvious  sum  rule
(\ref{eq:sumj1})  actually  implies  that the calculation 
of the partial  cross sections   requires unavoidably a 
complete  theory  for  the  inelastic  hadron--hadron cross section.

As  a  first step toward the calculation of the set of partial 
cross sections  $\sigma_k^{\rm jet}$ one can  note  
that it is  natural  to expect  that  
collisions  at different  impact parameter
will  result in a  different  number of  hard interactions, with small (large) 
$b$    corresponding  to a  larger  (smaller) number of interactions.
The {\em  average} number of  interactions at a fixed  impact parameter
$b$ can  be calculated as:
\begin{equation}
\begin{aligned}
\langle  n_{\rm jet} (b,s, p_\perp^{\rm min}) \rangle  &=
\int d^2b_1 ~\int d^2b_2 ~
P_{\rm int} (\vec{b} - \vec{b}_1 + \vec{b}_2) ~ \times ~
\\
~~~~~~~~~~~~~~& 
 \int dp_\perp 
~\int dx_1
~\int dx_2
~ \sum_{j, k, j^\prime, k^\prime} ~
~F^{h_1}_j (x_1,  b_1, \mu^2) 
~
F^{h_2}_k (x_2, b_2, \mu^2) 
~\frac{d \hat{\sigma}_{jk\to j^\prime k^\prime}}{dp_\perp} (p_\perp, \hat{s})
\end{aligned}
\label{eq:jet4}
\end{equation}
(where we  have  left implicit  the 
integration limits over  $p_\perp$, $x_1$ and $x_2$).
%that are identical to   equation (\ref{eq:jet2a})).
The expression  (\ref{eq:jet4})   differs from equation (\ref{eq:jet2a})
for  three  reasons:  (i) it replaces the   PDF's  $f_j^h(x, \mu^2)$   with 
the impact parameter   dependent   PDF's  $F_j^h(x, b, \mu^2)$;
(ii)   it includes two  additional  integrations 
over   $\vec{b}_1$ and  $\vec{b}_2$  that  describe 
 the positions  in transverse space
of  the partons inside  the two  hadrons; (iii) the function 
$P_{\rm int} (\vec{b}-\vec{b}_1 + \vec{b}_2)$  is also  included.

The impact parameter dependent  PDF's
$F^{h}_j (x,  b, \mu^2)$ 
describe  the probability to  find a  parton of type $j$ in hadron
$h$ with fractional longitudinal momentum  $x$ at transverse position
$b$ (with respect to the hadron c.m.).
These  functions   are  related to the standard PDF's by the 
relation: 
\begin{equation}
f_j^{h} (x, \mu^2) = \int d^2 b ~F^{h}_j (x,  b, \mu^2) 
\label{eq:Fnorm}
\end{equation}

The function $P_{\rm int} (\vec{b})$   describe  the 
probability   density that  two partons  
(each  in  a  different  hadron) separated by the  distance $b$ 
in   transverse space interact. The  function has the  normalization:
\begin{equation}
\int d^2 b ~P_{\rm int} (\vec{b}) = 1
\label{eq:Pnorm}
\end{equation}  
The  simplest   choice for    $P_{\rm int}(\vec{b})$, 
is  a delta  function
($P_{\rm int}(\vec{b}) = \delta^{2}(\vec{b})$),
or more  in general  (but at the cost of
introducing a new  parameter) 
a  gaussian  of  width $\sigma$.

Using  equations (\ref{eq:Fnorm}) and (\ref{eq:Pnorm}) 
 it is  simple to  verify that 
integrating $\langle  n_{\rm jet} (b,s) \rangle$ 
over all  impact
parameters  one  obtains  the  correct result: 
\begin{equation}
\int d^2 b ~ \langle n_{\rm jet} (b,s) \rangle 
= \sigma_{\rm jet} (s)
\end{equation}

The  physical  interpretation of   equation (\ref{eq:jet4})  
is  very straightforward.  To  obtain the  expected number  of hard  scattering
in  a  hadron  collision at impact parameter  $b$ 
one must integrate over the   distribution   of the partons 
in  trasverse  space  around  the  center of mass of the  colliding hadrons.
The  standard  PDF's   integrate over   this  transverse  space
variables, and are therefore  insufficient   for the   calculation, 
and are only capable  of  giving the inclusive  result.

To proceed in the  calculation one  must  obtain informations  about the
impact parameter  dependent  PDF's.
The simplest hypothesis  is to assume
that  the  dependence on  impact parameter 
of the $F_j^h$ functions can be factorized:
\begin{equation}
F_j^h (x, b, \mu^2) = f^h_j  (x, \mu^2) \; \hat{\rho}_h (b) 
\label{eq:factorization}
\end{equation}
where the  function $\hat{\rho}_h(b)$ satisfies the normalization condition:
\begin{equation}
\int d^2 b~ \hat{\rho}_h (b) = 1.
\label{eq:rhonorm1}
\end{equation}
With this factorization assumption   the 
quantity $\langle n(b,s)\rangle$   can be  written as:
\begin{equation}
\langle n_{\rm jet} (b,s)  \rangle = \sigma_{\rm jet} (s)
 ~ A(b)
\label{eq:factorization1}
\end{equation}
where the ``hadron overlap function''  $A(b)$ is:
\begin{equation}
A(b)  = \int d^2b_1 ~\int d^2b_2 ~\hat{\rho}_{h_1}(b_1) ~\hat{\rho}_{h_2} (b_2)
~P_{\rm int} (\vec{b} - \vec{b}_1 + \vec{b}_2)
\end{equation}
and  (using (\ref{eq:Pnorm}) and (\ref{eq:rhonorm1}))
satisfies the normalization condition:
\begin{equation}
\int d^2b ~A(b) = 1
\label{eq:anorm}
\end{equation}

A  reasonable  first approximation for  the overlap  function
$A(b)$  proposed  by Durand and Pi \cite{Durand:1988cr}  is to 
estimate it from the electromagnetic   form  factor of the colliding hadrons
(see appendix~\ref{sec:overlap}). Another (more phenomenological) approach has  been to
assume  a gaussian form, of a superposition of Gaussians.

It  should however be  stressed  that  the  factorization hypothesis  
(\ref{eq:factorization}) has no  serious
motivation beyond its simplicity, and it is likely to 
be  incorrect.  
In fact, the impact parameter PDF's  can  be  in principle
calculated  from the  generalized parton distribution functions (GPDF's)  
\cite{Burkardt:2002hr,Diehl:2003ny,Belitsky:2005qn}. 
Work on this  subject is in progress.

The need  to consider the dependence  on the transverse  degrees of  freedom  
introduces  a  serious  complication  and uncertainty in the 
calculation of the partial   cross  sections $\sigma_k^{\rm jet}$, but
unfortunately, even   having a good  theoretical  control of the overlap  
function  $A(b)$ (or  better   a detailed  knowledge  of the impact parameter
dependent PDF's)  is  not sufficient to  estimate
the partial  cross sections.

To complete  the calculation one  must make some  hypothesis
about  the fluctuations  in the number of  hard  interactions
for  collisions at a fixed  impact parameter  (and c.m. energy).
The simplest hypothesis is  to assume that 
the fluctuations  are simply  Poissonian.
The   partial  jet cross sections are then  calculable
integrating over  all impact parameters   the Poisson  probability:
\begin{equation}
\sigma_k^{\rm jet}  (s) 
 = \int d^2b ~\frac{\langle n_{\rm jet} (b,s)\rangle^k}{k!}  
~e^{-\langle n_{\rm jet}(b,s)\rangle}
\label{eq:pjet0}
\end{equation}

The hypothesis that the  multiplicity distribution of  the  hard interactions
at  a  given impact parameter  is Poissonian  is  however  not  necessarily correct,
and some simple  considerations  suggest  that 
in fact this distribution is  considerably  broader  than a Poissonian.
An  argument  in this  direction  can be developed as  follows.
At the instant $t$ a  hadron can  be  described   as  an  ensemble 
of  partons  each  having   a certain longitudinal momentum $x$   and
a trasverse position $\vec{b}$.
The set of  values
$\{q_j, x_j, \vec{b}_j\}_{(j=1,N)}$  
($q_j$ is  the complete set of  quantum  numbers of each parton)
is the   ``instantaneous  configuration'' 
of   the  hadron and  will be  denoted  
with the symbol ${\mathbb{C}}$.
The  probability to  find  hadron  $h$  in  a  certain instantaneous 
 configuration ${\mathbb{C}}$ can  be  denoted as  $P_h ({\mathbb{C}})$    with  the normalization
condition:
\begin{equation}
\int d{\mathbb{C}} ~P_h({\mathbb{C}}) = 1 
\label{eq:ccnorm}
\end{equation}
where the  integration  over $d{\mathbb{C}}$ indicates formally the sum over all
possible  configurations.

The interaction between  (for example)  two protons, in the  c.m.  frame,  lasts 
a crossing time  $t_{\rm cross}$, that  can  be estimated
as:
\begin{equation}
t_{\rm cross} \sim \frac{2 \, R_p}{\gamma_{\rm cm}}  \simeq
\frac{4 \, R_p \, m_p}{\sqrt{s}}
\end{equation}
where $R_p$ is the  linear  size of the proton  of order 0.5~fm.
Because of the Lorentz  length contraction  the  crossing
time  shrinks $\propto s^{-1/2}$

The time  required    for  the  radiation and  absorption of
partons, that is for  changing the parton  configurations 
is of order $R_p$  and is  therefore  much longer than the interaction time.
It is  therefore  a reasonable  approximation to  assume  that during  one interaction
the  hadrons  appear as   ``frozen''   each into its  own    configuration.

The expected  number of  hard interactions  in a  collision of impact parameter
$b$  is  determined  by the  parton  configurations   of the two colliding  hadrons.
If the configuration  is  composed by   few  hard partons, as   for example is the case
when most of the hadron energy is carried  by the valence  quarks,
the number of  hard  scattering  is  suppressed, while if the configurations
contain many  gluons with fractional energy of order
$x \sim (p_\perp^{\rm min})^2/s$  the number of hard interactions is  enhanced.

We  will   denote  as
$n_{\rm jet}(b, {\mathbb{C}_1},{\mathbb{C}_2})$ the  expected 
number of hard interactions in a collision
at impact parameter  $b$ between two 
hadrons with   configurations  
 ${\mathbb{C}_1}$
and 
 ${\mathbb{C}_2}$.
Integrating  over all  possible  hadron  configurations  one must
recover  the expected   value  of  the jet  multiplicity, therefore
we can write formally:
\begin{equation}
 ~\int d{\mathbb{C}_1}
 ~\int d{\mathbb{C}_2} ~ 
P_{h_1} ({\mathbb{C}_1})
~P_{h_2} ({\mathbb{C}_2})~
 n_{\rm jet}(b, {\mathbb{C}_1},{\mathbb{C}_2})
= \langle n_{\rm jet}(b,s) \rangle
\label{eq:conf0} ~.
\end{equation}

If we make the   further  assumption  that   the  actual 
number of  interactions  in a single  collision
at  impact parameter  $b$ with the   hadrons 
in  the configurations  ${\mathbb{C}_1}$ and ${\mathbb{C}_2}$
has  a Poisson distribution around the   expected  value,
we can now  write the 
partial jet cross sections as:
\begin{equation}
\sigma_k^{\rm jet} = \int d^2b
 ~\int d{\mathbb{C}_1}
 ~\int d{\mathbb{C}_2} ~ 
P_{h_1} ({\mathbb{C}_1})
~P_{h_2} ({\mathbb{C}_2})~
\left \{ \frac{\left [ n_{\rm jet}(b, {\mathbb{C}_1},{\mathbb{C}_2})\right ]^k}{k!} 
~\exp \left [ -n_{\rm jet}(b, {\mathbb{C}_1},{\mathbb{C}_2}) \right ]
\right \}
\label{eq:partial1}
\end{equation}

Of course equation 
(\ref{eq:partial1}) is  only a formal solution of  our problem
since  we have  not yet developed the
instruments to  
estimate  the probability of the  different
configurations and to perform the 
integration over the configurations space.

In order to  make progress we  will  make the 
assumption  that the expected  value of  the jet multiplicity
for  a certain  configuration of the hadrons is related 
to the averaged one   by the factorized  relation:
\begin{equation}
 n_{\rm jet}(b, {\mathbb{C}_1},{\mathbb{C}_2})  =
\langle n_{\rm jet} (b,s) \rangle ~
\alpha({\mathbb{C}_1}, {\mathbb{C}_2})
\end{equation}
where $\alpha({\mathbb{C}_1}, {\mathbb{C}_2})$ is  a  real  positive number.
With this  simplifying  assumption   we  can construct
a  function $p(\alpha)$ that is  independent  from $b$:
\begin{equation}
p(\alpha) = 
  \int d{\mathbb{C}_1} 
 ~\int d{\mathbb{C}_2} ~ P_{h_1} ({\mathbb{C}_1}) ~P_{h_2} ({\mathbb{C}_2})~
\delta \left [ 
\alpha({\mathbb{C}_1}, {\mathbb{C}_2})
- \alpha 
\right ] 
\label{eq:palpha1}
\end{equation}
It is  straightforward to see  that  because of the  normalization condition
(\ref{eq:ccnorm}) one  has:
\begin{equation}
\int_0^\infty d\alpha ~p(\alpha) = 1
\end{equation}
consistently  with  the  interpretation   of $p(\alpha)$ as a probability  density;
moreover because of equation (\ref{eq:conf0})  one has  also that:
\begin{equation}
\int_0^\infty d\alpha ~\alpha~p(\alpha) = 1
\label{eq:palpha2}
\end{equation}

Using the  definition  of (\ref{eq:palpha1}) one  can rewrite  
(\ref{eq:partial1}) in the simpler form:
\begin{equation}
\sigma_k^{\rm jet} = \int d^2b
 ~\int_0^\infty d\alpha ~p(\alpha)
\left \{ \frac{\alpha^k \; \langle n_{\rm jet}(b,s) \rangle^k}{k!} 
~\exp \left [ -\alpha \; \langle n_{\rm jet}(b, s) \rangle \right ]
\right \}
\label{eq:partial2}
\end{equation}
In this equation we have  finally arrived  to  write
the  jet multiplicity distribution in terms  of 
 $\langle  n_{\rm jet}(b,s) \rangle$  and
the function  $p(\alpha)$, that is  unknown,  but  has  
fixed  normalization  and  first  moment.
The expression  for the 
partial  jet cross sections  given  in equation (\ref{eq:pjet0})) 
that  was calculated ignoring  the effects
of  fluctuations in the parton  configurations 
%assuming simple Poissonian  fluctuations  at  each impact   parameter
%$b$ (according to  equation (\ref{eq:pjet0})) 
is recovered in the limit  where  the function  $p(\alpha)$   has  
vanishing  width  and reduces 
to the form  $p(\alpha) = \delta [\alpha -1]$.

It is  straightforward  to see   (using equation (\ref{eq:palpha2}))
that one  has:
$\langle k \rangle = \sigma_{\rm jet}/\sigma_{\rm inel}$ in  agreement
with equation (\ref{eq:kmed}).
The second  moment of the  jet  multiplicity
distribution can  be expressed as:
\begin{equation}
\langle k^2 \rangle = 
\frac{1}{\sigma_{\rm inel}} \; \sum_k  k^2 \; \sigma_k^{\rm jet} = 
\frac{\sigma_{\rm jet}}{\sigma_{\rm inel}}
+ \frac{(1 + w)}{\sigma_{\rm inel}} 
~\int d^2b ~\langle n_{\rm jet} (b,s) \rangle^2
\label{eq:wwa}
\end{equation}
where 
\begin{equation}
w = \int_0^\infty d\alpha~(\alpha^2 - 1)~p(\alpha)
\end{equation}
is  the variance  of the $p(\alpha)$  distribution.
Using the factorization  hypothesis 
(\ref{eq:factorization1})  
one can rewrite equation (\ref{eq:wwa}) as:
\begin{equation}
\langle k^2 \rangle = \frac{\sigma_{\rm jet}}{\sigma_{\rm inel}}
+ (1 + w)~\frac{\sigma_{\rm jet}^2}{\sigma_{\rm inel}} 
~\int d^2b ~[A(b)]^2 
\label{eq:wwb}
\end{equation}
The  width of the jet  multiplicity  distribution  is therefore
determined  by  the geometry of  the hadronic  matter  (that is  from
the shape  of the   overlap  function $A(b)$) {\em and}  from the 
variance $w$  of   the function $p(\alpha)$.
Note  that  the  simple  eikonal model  corresponds to  a  vanishing  $w$
and  therefore  to  the minimum   possible  $\langle  k^2 \rangle$
(for a fixed  overlap  function $A(b)$).

The partial  cross sections  $\sigma_k^{\rm jet}$ 
and the quantity $\langle k^2\rangle$ are
(at least in principle)  measurable with
observations of the jet multiplicity  distribution.
From these measurements one can obtain information about the properties
of  the function $p(\alpha)$.

In fact we  have  already  some  (indirect) 
information about the  jet   multiplicity
distribution, and  there are  indications  that predictions
based on the simple eikonal approach are not adequate.
Some of the most  sophisticated  Montecarlo instruments  
for the study of  high energy  hadron  collisions, such as
PYTHIA  \cite{pythia0,pythia}  and  HERWIG  \cite{herwig,herwig1} 
include  a  treatment of 
multi--parton interactions following the simple eikonal model.
The  algorithms  for  multi--parton interactions 
in these Montecarlo  codes compute an  inclusive  jet cross
section $\sigma_{\rm jet}$, assume an overlap  function $A(b)$,
and  them generate  a number of  elementary  interactions according to
the probability  distribution of   equation (\ref{eq:pjet0})
or  equivalently  of  equation (\ref{eq:partial2}) with 
$p(\alpha) = \delta [\alpha -1]$.
In most cases the presence  of multiple  interactions  cannot be
experimentally resolved  but  is  detectable   statistically, for example 
from fluctuations in the  charged particles   multiplicity distribution.
In order to  reproduce  the broad fluctuations of the data,  it appears   necessary
to   construct  some   {\em ad hoc}  functional form
for  the overlap  function.  For example PYTHIA \cite{pythia0,pythia}  uses
a  double--Gaussian  with a  denser core and more extended halo.
The  overlap  function $A(b)$  however (as  will be discussed  in more detail in
the following)  in principle also
determines   the  $t$  distribution of elastic  scattering,
and therefore one does  not have the freedom to  modify it in  an  arbitrary way.

The introduction  of  a  non vanishing variance for the function  $p(\alpha)$
allows  to  modify the  width of the jet multiplicity distribution
without  modifications  of the overlap  function  $A(b)$.

One open problem in high energy hadron  collisions, 
that  is likely to  be  relevant in the interpretation
of future data on possible   manifestations  of new  physics
in  high energy   collisions  is the problem  of the so  called
``underlying event''.
The  observations  \cite{Acosta:2004wqa,Affolder:2001xt} 
show  that  in   events  with the  presence of a high $p_\perp$ scattering, 
the ``environment''  that accompany  the  observed  jets 
has an average transverse momentum  higher 
than what is found in minimum bias  events.   
Montecarlo codes like HERWIG,  ISAJET or PYTHIA
at  present describe only partially  these  effects
 \cite{Acosta:2004wqa,Affolder:2001xt}.
The  origin of a higher ``ambient level''  of
$p_\perp$ for the underlying event can be related to the presence
of additional (softer and unresolved) parton scatterings
that accompany the observed jets.
The correct description of the  underlying event is therefore 
 related  to a good theoretical  control of the   multiplicity 
distribution of  parton interactions.
A  Montecarlo  inplementation of  multi--parton interactions
based  on an impact parameter   picture  
for hadronic  collisions naturally  contains  
some of the qualitative features
observed in  underlying events,  because events selected with 
a  high $p_\perp$ jets  are more  likely to be  central collisions,
and  therefore  are more likely to  contain additional
parton  scatterings.
A  non vanishing  variance of $p(\alpha)$
should   however  
enhance the differences   between   underlying  and  minimum bias events,
and could  therefore  play a  non  negligible
role in  the description of the data.
A  quantitative  study of this  problem 
with  Montecarlo methods  is  a goal  for future  work.

The  problem of the fluctuations in the configurations of 
partons  in a  hadron  is  also   intimately related to
diffractive   scattering, as  we  will   illustrate in the  following
section.

\section{Total  and  elastic cross sections}
\label{sec:total}
\subsection{General Formalism}
\label{sec:general}
In the following we  will  need to consider  also the elastic  and
total  cross section  in hadronic  scattering.
The elastic  scattering amplitude
in the  collision  of  two  hadrons  of  type $h_1$ and $h_2$ 
at  c.m. energy $\sqrt{s}$  can  be  written \cite{Block:1984ru}
 as  a (2--dimensional) integral  over  impact parameter:
\begin{equation}
F_{\rm el} (q,s) 
 = i\; \int \frac{d^2b}{2 \, \pi} ~ e^{i \vec{q}.\vec{b}}~
\Gamma_{\rm el} (b, s)
\label{eq:gen_ampl}
\end{equation}
where $\Gamma_{\rm el} (b,s)$ is  the  profile function, that without loss of 
generality can be  written as:
\begin{equation}
\Gamma_{\rm el} (b,s) = 1- e^{-\chi(b,s)} 
\end{equation}
with  $\chi(b,s)$ the eikonal function.
In the notation we are  leaving  implicit the  dependence
on the type of hadrons  participating in the collision.
The  elastic scattering amplitude   
is  related to the differential cross section   by:
\begin{equation}
\frac{d \sigma_{\rm el}}{dt} (t, s) = 
\pi \; \frac{d \sigma_{\rm el}}{d^2 q} (\vec{q}, s) = 
\pi \;
\left | F_{\rm el} (\sqrt{-t}, s) \right |^2
\label{eq:gen_dsig}
\end{equation}
In this  equation  $t = (p_{h_1} - p_{h_1}^\prime)^2$ is 
the squared momentum transfer, and we have   used the  approximation 
$t = -|\vec{q}|^2$
that  is  valid for small $-t$ and high energy.
Integrating over all $t$ 
 one obtains:
\begin{equation}
\sigma_{\rm el}(s)  = 
 \int d^2b ~\left | \Gamma_{\rm el} (b,s) \right |^2 
\label{eq:siggen_el}
\end{equation}
The total  cross section is related to the  imaginary part of the
forward  elastic  scattering  amplitude by the optical 
theorem  and is  given by:
\begin{equation}
\sigma_{\rm tot}(s)  =  
4 \, \pi \; \Im [F_{\rm el} (0, s) ] = 
2 \;\int d^2b ~ \Re \left [ \Gamma_{\rm el} (b,s) \right ] 
\label{eq:siggen_tot}
\end{equation}
Combining  equations 
(\ref{eq:siggen_el})  and
(\ref{eq:siggen_tot}) 
one obtains an expression 
for the inelastic cross section as an integral over the impact parameter:
\begin{equation}
\sigma_{\rm inel}(s)  =  \int d^2b ~\left \{1 - \left |1 - \Gamma_{\rm el} (b,s) \right |^2
 \right \} 
\label{eq:siggen_inel}
\end{equation}

Equation (\ref{eq:gen_dsig})
gives  the  exact shape of the 
differential elastic cross section.  For small $|t|$  this
shape  is  to  a good approximation a simple
exponential   ($d\sigma_{\rm el}/dt \propto e^{B \, t}$)
and it is   convenient
to   define  the slope   $B_{\rm el}$ of  elastic scattering:
\begin{equation}
B_{\rm el}(s) = \left [ \left (\frac{d\sigma_{\rm el}}{dt} ~\right )^{-1}
\frac{d}{dt} \left ( \frac{d \sigma_{\rm el}}{dt} \right ) \right ]_{t=0} 
\end{equation}
It is  straightforward to see  that   if  the 
profile  function is real,   the slope $B_{\rm el}$  
 can be calculated as:
\begin{equation}
B_{\rm el}(s) = 
\left \{ \int d^2b ~ \frac{b^2}{2} ~ \Gamma_{\rm el} (b,s) 
\right \}
\times 
\left \{ \int d^2b ~  
\Gamma_{\rm el} (b,s)   \right  \}^{-1} = 
\frac{\langle b^2 \rangle } {2}
\label{eq:Bdef}
\end{equation}
and measures  the  value   $\langle b^2  \rangle$,
of the profile  function.

Using the approximate    exponential  shape  of the differential  cross section one 
finds the relation
\begin{equation}
\sigma_{\rm el} (s)  \simeq \pi ~ \frac{\left | F_{\rm el} (0,s) \right |^2}
{B_{\rm el}(s)}
= \pi \; \frac{\left (\Im \left [ F_{\rm el} (0,s) \right ] \right )^2 
~\left | i + \rho \right |^2}{B_{\rm el}(s)}
\end{equation}
(where $\rho$ is the ratio of the real to  imaginary parts of the
forward  elastic amplitude).
From the optical theorem (\ref{eq:siggen_tot})  one  obtains
the relation:
\begin{equation}
\sigma_{\rm el} = \frac{\sigma_{\rm tot}^2 \; (1 + \rho^2) }{16 \, \pi \, B_{\rm el}}
\label{eq:unit1}
\end{equation}

\subsection{The simple  eikonal  model}
\label{sec:model0}
The  expressions   (\ref{eq:gen_ampl}--\ref{eq:siggen_inel}) 
allow to  compute  the total and elastic cross sections
for   hadron collisions
in terms  of the  profile  function $\Gamma_{\rm el} (b, s)$  or equivalently
of the  eikonal  function $\chi(b,s)$.
We  remain with the task  of  constructing these  functions.  
Physical insight on  $\Gamma_{\rm el} (b,s)$  and $\chi(b,s)$
can be obtained  using a 
well known   analogy with the  classical  treatment  
of the absorption and scattering of  a  plane  wave  of  light  
from  an  opaque   screen.
If the  ratio  between 
the amplitude ``just behind''  the screen 
and   the amplitude of the  incident  plane wave is  $\Gamma(\vec{b})$
(where  $\vec{b}$ is a 2--dimensional vector  spanning the   screen),
then  it is simple to   obtain expressions  for the 
total, elastic and absorption    cross sections  that are
formally  identical to  equations
(\ref{eq:gen_ampl}--\ref{eq:siggen_inel}). 
In particular  the  expression  (\ref{eq:siggen_inel}) 
for the  inelastic  cross section 
suggests to interpret  the quantity $1-| 1- \Gamma (\vec{b})|^2$  as 
the probability   to  absorb the wave 
at  the  position $\vec{b}$ on the screen.

At high energy the   elastic  scattering amplitude, in   reasonably 
good approximation,  is purely imaginary, and accordingly 
the profile  and  eikonal functions  are  purely real.
Neglecting the real  part  of the scattering amplitude
the  optical  analogy  is  then sufficient to express  the 
profile  function as:
\begin{equation}
\Gamma_{\rm el} (b, s) = 1 - \sqrt{1 - P_{\rm abs} (b,s)}
\label{eq:ansatz}
\end{equation}
with  $P_{\rm abs} (b,s)$ the absorption probability.

In the following we will use the  approximation 
to  consider the elastic scattering  amplitude as  purely imaginary,
and the profile and eikonal functions as purely real.
The  analiticity of the elastic scattering amplitude that is
necessary to respect causality,  can be imposed estimating the
real part  with a dispersion relation.

Equation (\ref{eq:ansatz})  can  be re--expressed in the form:
\begin{equation}
\Gamma_{\rm el} (b, s) \equiv
1 - e^{-\chi(b,s)} =
1 - \sqrt{P_0(b,s)}
=  1 - \exp \left [ - \frac{\langle n(b,s) \rangle }{2} \right ]
\label{eq:ansatz1}
\end{equation}
In this  equation  $P_0 = 1 -P_{\rm abs} = e^{-\langle n (b,s)\rangle}$ 
is  the  probability of {\em no}   absorption  in a    collision   at a certain $b$ and $s$.
The  notation  $P_0(b,s) = e^{-\langle n(b,s) \rangle}$ is  motivated by  the
physical  ansatz  that ``absorption'' 
in a hadron collision  corresponds to  
at least one  interaction    between  the parton  constituents of the two hadrons.
If   $\langle  n(b,s) \rangle$ is the  expected  number of  {\em all}  
elementary interactions  between partons,  and if the fluctuations
of this average  number of  interactions   is Poissonian, 
one   obtains  equation (\ref{eq:ansatz1})  for the  profile  and eikonal  function.

The ansatz  outlined  above  connects  the profile  function
$\Gamma_{\rm el}(b,s)$ to $\langle n(b,s)\rangle$
that is the    average number  of  elementary interactions
at  impact parameter $b$ and  c.m. energy $\sqrt{s}$.
In the  following we will call this model
as the ``simple  eikonal model''.
This ansatz   was introduced 
by  Durand and Pi in  \cite{Durand:1988cr},
who  proposed to compute  the quantity $\langle  n(b,s) \rangle$ as:
\begin{equation}
\langle  n(b,s) \rangle  = \sigma_{\rm jet} (s, p_\perp^{\rm min}) ~ A(b)
\end{equation}
where   $\sigma_{\rm jet} (s, p_\perp^{\rm min})$ is  
the quantity  discussed in the previous  section, that is the inclusive cross section
for the  production 
of  jet pairs  above   a certain (fixed) $p_\perp$, and  
the   energy independent geometry   factor  $A(b)$ 
gives the  overlap  of  hadronic  matter.
For $pp$ interactions   Durand and Pi  (see  discussion in appendix~\ref{sec:overlap})
chose:
\begin{equation}
A_{pp} (b) = \frac{b^3}{96 \, \pi \; R_p^5}  \; K_3 \left ( \frac{ b}{R_p} \right )
\label{eq:App}
\end{equation}
with  $R_p^{-2} = 0.71$~GeV$^2$. 

The model  in this simplest  form (that  has  in fact  a single parameter, the value 
of  $p_\perp^{\rm min}$) soon proved to be inconsistent  with the data.
Perhaps  the main  difficulty with the  original  formulation
of the model  is that it  predicts  
an  incorrect  relation between  $\sigma_{\rm  tot}(s)$ and  $B_{\rm el}(s)$.
This problem  in  fact  is  of a general  nature
and  indicates that  in the simple  eikonal model, the  factorization
of  $\langle  n(b, s)\rangle $  as  the product  of two
functions  one of  $s$ and the other  of $b$  cannot  fit the data.  

To falsify the factorization hypothesis 
 (\ref{eq:factorization1})   it is in fact sufficient
to measure  both  $\sigma_{\rm tot}$ and  the  slope of elastic scattering $B_{\rm el}$
(or more in general,  because of   equation (\ref{eq:unit1})
and the smallness of $\rho$,  
two out of the three  quantities  $\sigma_{\rm tot}$, $\sigma_{\rm el}$ and $B_{\rm el}$)
at two  different  values of $\sqrt{s}$.
It is  always  possible  to  find 
a  profile  function $A(b)$ and  a  value of the   eikonal  function
$\sigma_{\rm  eik}(s_1)$    
  that   reproduce the observations
of  $\sigma_{\rm tot}$ and $B_{\rm el}$  at   c.m. energy
$\sqrt{s_1}$, but  then  one  remains  with
a  single parameter  $\sigma_{\rm eik}(s_2)$  that can  be chosen to
reproduce $\sigma_{\rm tot}(s_2)$ or  $B_{\rm el}(s_2)$,  however  the
two  values will coincide only if  the factorization
hypothesis  is valid.

Models  based on an eikonal  model  and the 
factorization hypothesis  (\ref{eq:factorization1})   naturally
predict  a correlation  in the growth  of $\sigma_{\rm tot} (s)$  and
$B_{\rm el}(s)$  with energy,   however   the 
growth of $B_{\rm el}(s)$ in the data is faster than  the predictions,
as  was  realyzed very early  by Durand  and  Pi  themselves \cite{Durand:1988ax}.

\begin{figure}[hbt]
\begin{center}
 \includegraphics[angle=90.,width=14.cm]{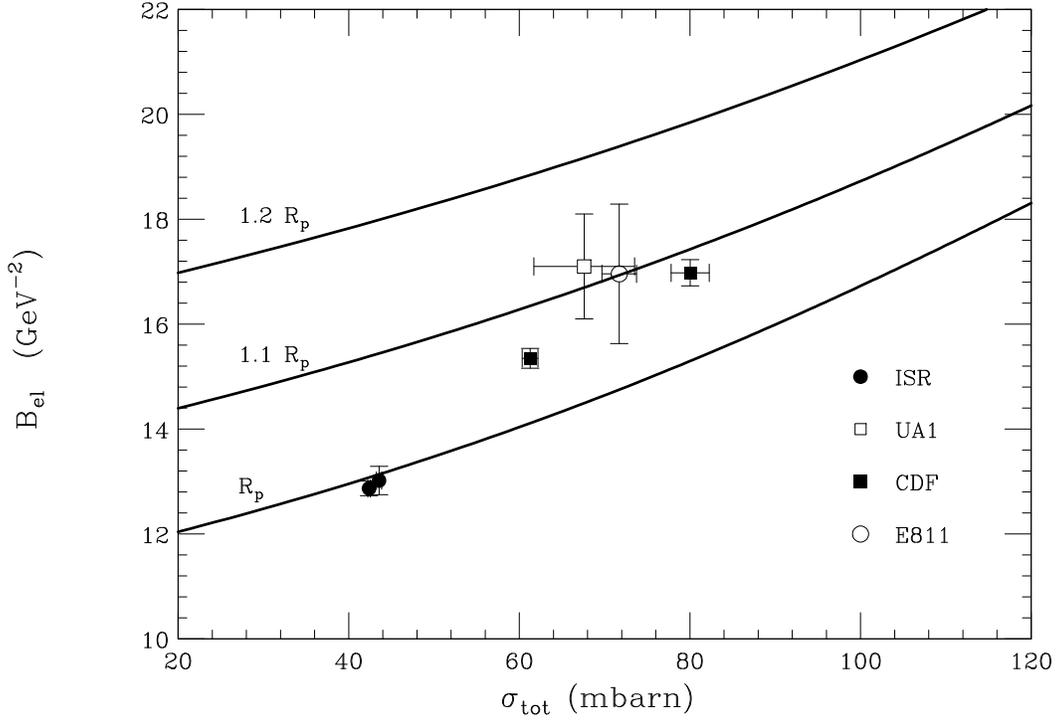}
\end{center}
   \caption{\small  
The points  are measurements 
of  the total  cross section $\sigma_{\rm tot}$ and of 
the forward  slope  $B_{\rm el}$ of the  elastic  scattering
for  $pp$ and   $\overline{p}p$ collisions at collider energies.
The lines  correspond to  predictions  
based  on the simple eikonal model  using  the parametrization
of  equation (\protect\ref{eq:pareik})
for  $\langle n(b,s) \rangle$.
The three lines  are  computed for three  values of
the $r_0$ parameter 
($r_0 = R_p$, 1.1~$R_p$ and $1.2~R_p$).
The  ISR data at  $\sqrt{s} =  52.8$ and 62.3~Gev  is from   \protect\cite{Amos:1985wx};
 the UA1  data at $\sqrt{s} = 540$~GeV   from
 \protect\cite{Arnison:1983mm};  the CDF data  at $\sqrt{s} = 546$ and 1800~GeV from
 \protect\cite{Abe:1993xy,Abe:1993xx}; 
the E811 data  at 1800~GeV   from  \protect\cite{Avila:1998ej}.
\label{fig:bsig}  }
\end{figure}

This problem  is  illustrated  in  fig.~\ref{fig:bsig}
where we  show,  in the plane $(\sigma_{\rm tot}, B)$,
experimental data  obtained 
for  $pp$   scattering at the ISR \cite{Amos:1985wx}
and   for  $p\overline{p}$  scattering at  CERN  \cite{Arnison:1983mm}
and  Fermilab \cite{Avila:1998ej,Abe:1993xy,Abe:1993xx},
and  compare with the predictions  of the  simple  eikonal  model.
The thick  curve is the  prediction of 
the simple  eikonal  model      using  a 
factorized  form  of  $\langle n(b,s)\rangle$   and using  for  $A(b)$    the 
expression  of equation (\ref{eq:App}).  This line  passes through the
ISR  data  points, but fails  at higher energy.
The other  lines in the figure  are calculated using the same functional
form  for  $A(b)$  but  replacing  the parameter $R_p$ with
$r_0 = 1.1~R_p$ and $r_0 = 1.2~R_p$.
It is  clear  that, in the  simple eikonal model 
the  width  in impact parameter
of the function $\langle  n(b,s) \rangle$ cannot be  energy independent 
and must {\em  grow} with $s$  in order
to reproduce the data. 

An additional problem  for the  Durand and Pi model,
beside  the one  we have just discussed
connected  to the relation between $\sigma_{\rm tot}$ and $B_{\rm el}(s)$,   is  also
that the  energy dependence of $\sigma_{\rm tot} (s)$ of the model,
driven   by the growth  of $\sigma_{\rm jet} (s, p_\perp^{\rm min})$
with $s$ for $p_\perp^{\rm min}$  fixed is significantly faster  than the data.

The ansatz of the simple eikonal model has been considered 
by  several other authors \cite{DiasdeDeus:1987yw,
Margolis:1988ws,
Capella:1992yb,
Werner:1993uh,
Fletcher:1994bd,
Godbole:2004kx}, that have constructed
different models for the function $\langle n(b,s)\rangle$,
abandoning the factorization hypothesis.
For example,  it has   been suggested to decompose the
function $\langle n(b,s) \rangle$   in the general form:
\begin{equation}
\langle n(b,s) \rangle 
= \sigma_{\rm eik} (s) \; A(b,s) = \sigma_{\rm soft} (s) \; A_{\rm  soft} (b,s)
+  \sigma_{\rm hard} (s) \; A_{\rm  hard} (b,s) 
\end{equation}
where the  two   terms  describe a ``hard''  contribution
that can be calculated with    perturbative  methods, and
a  ``soft'' non perturbative part.
As discussed above, the 
width in impact parameter   of the combined
overlap function  $A(b,s)$  must increase with $s$   to reproduce
the relation between  $B_{\rm el}(s)$ and  $\sigma_{\rm tot}(s)$.
Several justifications  of this  growth have  been offered, however
in our opinion  none is  really convincing.

The  energy dependence  of  $\sigma_{\rm eik} (s)$, and its  decomposition
in a soft and a hard part  have also been the object  of considerable
discussion. 
In this  work we will  not  enter  in this discussion, because
our main purpose here is  to discuss the limitations of
the  simple eikonal model. In the simple eikonal model  the 
quantity $\langle n (b,s)\rangle$ determines  completely the profile
function, and therefore the  total and  elastic  cross sections.
We will also  consider the  quantity $\langle  n(b,s)\rangle$    
attributing  to it the same  physical meaning
that it has in   the simple  eikonal mode,   namely the  average number of  elementary
interactions at impact parameter  $b$ and c.m. energy $\sqrt{s}$;
however in order to  describe in a consistent  way  inelastic diffraction,  we will
propose a different  method to  connect  $\langle  n(b,s)\rangle$ with the profile
function and to the observable cross sections.

In the following  we will parametrize the 
function  $\langle  n(b,s) \rangle$    as:
\begin{equation}
\langle n(b,s) \rangle  = \sigma_{\rm eik}(s) ~
\left \{ \frac{b^3}{96 \, \pi \; [r_0(s)]^5} 
\; K_3 \left ( \frac{ b}{r_0(s)} \right ) \right \}
\label{eq:pareik}
\end{equation}
with  an overlap  function   that  has the same  form as in 
equation (\ref{eq:App}), but    with 
an $s$ dependence  obtained
substituting   for  $R_p$ an energy  dependent
parameter  $r_0(s)$:

The   simple  eikonal model   ansatz  outlined  above  
does  not  include a  treatment of  the  inelastic  diffraction processes.
This is a  serious  limitation, because
the  consistent  description 
of these processes  is  in fact essential.
This  problem is addressed in  the next section.

\section{Inelastic  diffraction}
\label{sec:diff}
Inelastic  diffraction   \cite{Amaldi:1976gr} produces three classes  of events:
beam, target and double diffraction.
In beam  (target)  diffraction the  beam (target)  particle  
is  excited   to a  higher mass state
of the same  quantum  numbers
(with the possible  exception of spin)
while  the  other  initial  state  hadron  remains unchanged.
In double  diffraction  both  colliding hadrons are excited 
into  higher mass states.
The three  processes can therefore  be   described as:
\begin{equation}
\begin{split}
&h_1 \, h_2 \to h_1^* \, h_2     ~~~~~({\rm Beam~Diffraction})\\
&h_1 \, h_2 \to h_1 \, h_2^*     ~~~~~({\rm Target~Diffraction})   \\
&h_1 \, h_2 \to h_1^* \, h_2^*   ~~~~~({\rm Double~Diffraction})
\end{split}
\end{equation}
The experimental  selection of  events  that belongs to these
three  classes  is  not  trivial, especially for the 
double diffraction ones.  These events  are  characterized  by a 
large  ``rapidity gap'' that  separates  particles
produced in the decay of the  two  excited  states $h_1^*$ and
$h_2^*$.
A complete description of beam  (target) 
diffraction   is  given by 
the double  differential  cross sections $d^2\sigma/(dM_1^2 \, dt)$
($d^2\sigma/(dM_2^2 \, dt)$)   that gives the probability to produce
an event   where the beam (target) particle is  excited to a state
of mass $M_1 > m_1$  ($M_2 > m_2$)  with  transfer momentum $t$.
The  double  diffraction cross section  is    described by the 
3--times differential cross section:
 $d^3\sigma/(dM_1^2 \, dM_2^2 \,  dt)$.
For  single diffraction one  has  a reasonably accurate 
picture of the  $M$  and $t$   dependence of the cross section.
The  $t$ dependence has   the  characteristic    exponential
behaviour of all  diffractive--like  processes,  
for the mass  dependence
the cross section     grows  very  quickly   above the  threshold
($M = m_p + m_\pi$ for proton  excitation),  oscillates  following
the  structure of the resonances 
with the quantum  numbers of  the
hadron in question and  then  falls  at  higher masses   roughly 
proportionally to $M^{-2}$.  
In the following we will  indicate as ``diffraction''
the  sum of these three classes  of inelastic  diffractive events.

We  take the point of   view  that any  parton--parton scattering implies an
exchange of color,  and therefore,
in the  treatment  of the cross  section  given above, 
all events with one or  more  parton--parton  interaction 
must be considered  as non--diffractive.  
In the simple eikonal model the total  cross section   is therefore decomposed into
an inelastic non--diffractive part (with at least one  parton--parton
scattering)  and  the elastic part, with no  room left for  diffraction.
This   lack of inclusion  of  inelastic  diffraction
 is  an important  conceptual problem for the simpe eikonal model.

\subsection{Good and Walker model}
\label{sec:good}
The   fundamental  idea for the description of    inelastic  diffraction
in  hadronic  collisions  has been  introduced  long ago 
 by Good and  Walker \cite{Good:1960ba}
  as an analogy with the scattering  of polarized
light on a  bi--refringent absorbing medium.
It is  well known    that   if a plane wave of light impinges on
an  absorbing screen  
with     grayness profile  $\Gamma (b)$   one  has
absorption ($\propto 1 -  |1-\Gamma (b)|^2$)
and  elastic  scattering  ( $\propto |\Gamma (b)|^2$)
with a  very  forward diffraction  pattern   that depends
on  the geometry of the screen.

If  the screen  absorbs  differently  the light polarization states,
in general an  incident beam   of a  given initial polarization
will  result in absorption, and in  scattered  light with  the same
(elastic scattering) and the orthogonal (inelastic  diffraction)
polarization state.
For a more explicit example, let us consider
an  incident  beam of light in  the linear
polarization  state $|x\rangle$, and an absorbing 
screen   that  has the grayness profile 
 $\Gamma_{x^\prime}(b)$ 
for  light with polarization $|x^\prime\rangle$
and  the grayness profile 
 $\Gamma_{y^\prime}(b)$ for  light with polarization $|y^\prime\rangle$
with
\begin{eqnarray*}
|x^\prime\rangle & = &  \cos \varphi \; |x \rangle +  
\sin \varphi \; |y \rangle  \\
 ~ & ~& \\
 |y^\prime\rangle & = &  -\sin \varphi \; |x \rangle +   
\cos \varphi \; |y \rangle
\end{eqnarray*}
It is  clear that one will  have an absorption  cross section
according $1 - |\Gamma_{x^\prime}(b)|^2 - |\Gamma_{y^\prime}(b)|^2$
and   after the  screen one  will find  scattered waves with both
the  $|x\rangle$ and
  $|y\rangle$  polarizations.
The cross section for 
elastic  scattering and inelastic  diffraction turn  out to be:
\begin{equation}
\sigma_x = 
\int d^2b ~  \left | 1 -
[1 - \Gamma_{x^\prime} (b)] \cos^2 \varphi
-[1 - \Gamma_{y^\prime} (b)] \sin^2 \varphi \right  |^2
\end{equation}
\begin{equation}
\sigma_y = 
\int d^2b ~  \cos^2 \varphi \; \sin^2 \varphi ~| \Gamma_{y^\prime}(b) -
\Gamma_{x^\prime}(b)|^2
\end{equation}
Note that  inelastic  diffraction  is non vanishing only
when  $ \Gamma_{x^\prime} \ne \Gamma_{y^\prime}$, and   
if $\varphi$  is not  a multiple  of $\pi/2$, that is if the 
 states $\{|x\rangle,|y\rangle\}$
do not coincide  with  the states
 $\{|x^\prime\rangle,|y^\prime\rangle\}$  that are the  eigenstates
of the  transmission across the screen.

It is straightforward  \cite{Good:1960ba} to   generalize  this 
elementary example.
We can  consider two complete sets of 
orthonormal states   $\{|\varphi_m\rangle\}$ 
and   $\{|\psi_j\rangle\}$.
The states    $|\varphi_m\rangle$  are  directly  observable,
with the label  $m$  describing   the invariant masses of the 
two final  hadronic  states   (in the forward and backward hemisphere)
and their  particle   content. 
Without loss of  generality we can   assume
that the state $|\varphi_1\rangle$  with  $m=1$ corresponds
to the initial  state   of  the scattering
(for example in case of a  $\pi p$ collisions to the state $|\pi p\rangle$).

The states $|\psi_j\rangle$  are  eigenstates of the scattering
matrix. That is,  defining as usual the $S$ matrix
as $S = I + i \, T$    one has:
\begin{equation}
T \; |\psi_j \rangle = t_j \; |\psi_j\rangle
\end{equation}

The relation between the two  orthonormal  bases is  given by:
\begin{eqnarray}
|\varphi_m \rangle & =  & \sum_j C_{m j} \, |\psi_j\rangle 
\\
|\psi_j \rangle & =  & \sum_m C_{m j}^* \, |\varphi_m\rangle 
\end{eqnarray}
It is  now  possible to  write the different  components 
of the cross sections  as  integrals  over the
impact parameter  dependence  of the $t_j(b)$.
The absorption cross section   is:
\begin{equation}
\sigma_{\rm abs} = 
\int d^2b ~\left [ 1 -
\sum_j \left | C_{1 j} \, \{1-t_j(b) \} \right |^2 \right ]
\label{eq:siga_abs}
\end{equation}
The cross section for    diffraction into the state $m$ is:
\begin{equation}
\sigma_m = 
\int d^2b ~\left | \sum_j  C_{m j}^* \, C_{1 j} \; t_j(b) \right |^2 
\end{equation}
The  elastic cross section   corresponds   to $\sigma_1$ and  therefore is
\begin{equation}
\sigma_{\rm el} = 
\int d^2b ~\left | \sum_j  |C_{1j}|^2  \; t_j(b) \right |^2 
\label{eq:siga_el}
\end{equation}
Summing over all states $m$  and using 
the orthonormality of the states  one  obtains:
\begin{equation}
\sigma_{\rm diff + el} =
\sum_m \sigma_m =  
\int d^2b  \sum_j  |C_{1j}|^2  \; \left |t_j(b) \right |^2 
\label{eq:siga_difft}
\end{equation}
The  inelastic  diffraction  cross section can be obtained
 subtracting  equation (\ref{eq:siga_el})  from (\ref{eq:siga_difft}).
The total cross  section  can be obtained
summing  the  elastic, diffractive  and absorption cross sections  obtaining the
result:
\begin{equation}
\sigma_{\rm tot} =
\int d^2b  \sum_j  |C_{1j}|^2  \; 2  \; \Re \left [t_j(b) \right ] 
\label{eq:siga_tot}
\end{equation}
in agreement also with the optical theorem.

The problem  with the  approach  that  we have just outlined, is that it is only formal,
and  requires  the introduction of  many, in fact infinite  parameters;  moreover the nature
of the   eigenstates   $|\psi_j\rangle$ is   not  physically  obvious.

\subsection{Partonic  Interpretation  of the scattering  eigenstates}
\label{sec:partonic}
Miettinen  and  Pumplin 
\cite{Miettinen:1978jb}  have  made
the proposal to interpret  the  states     $|\psi_j\rangle$  
(the eigenstates  of the  $T$ matrix) as  parton configuration  states.
In their  paper they also  include  a simple  explicit  methods
to construct these partonic  models
(see \cite{Sapeta:2004av,Sapeta:2005ba}  for a recent  rediscussion).

In this work  we will develop the idea  that 
one needs to  integrate over the parton configurations
of the interacting  hadron.
The discussion  that we have outlined in  section~\ref{sec:multiple} 
can also be  applied  to the  current  problem.
Using the  notations  that we have introduced  in section~\ref{sec:multiple}
the   label $j$ of the    $T$ matrix  eigenstates  $|\psi_j\rangle$  
corresponds to the  direct product of the    configurations  ${\mathbb{C}_1}$ and  ${\mathbb{C}_2}$ of
the  colliding   hadrons, and the  squared coefficient  $|C_{1j}|^2$ 
corresponds  to the probability $P_{h_1}({\mathbb{C}_1}) \times P_{h_2} ({\mathbb{C}_2})$ 
to find  the two  hadrons  in a certain  configuration.
One  then has  also   the  correspondence:
$$
\sum_j \left | C_{1 j} \right |^2  \leftrightarrow  
\int d{\mathbb{C}_1} ~\int d{\mathbb{C}_2} ~ 
P_{h_1} ({\mathbb{C}_1})
~P_{h_2} ({\mathbb{C}_2})~
$$
The transmission  eigenvalues  $t_j(b)$ have   the  partonic  interpretation as:
\begin{equation}
t_j (b) = 1 - \exp \left [ - \frac{n_j (b)}{2} \right ] =
 1 - \exp \left [ - \frac{n (b,{\mathbb{C}_1},{\mathbb{C}_2})}{2} \right ]\
\label{eq:ident1}
\end{equation}
where as in section~\ref{sec:multiple}    $n(b,{\mathbb{C}_1},{\mathbb{C}_2})$ is the expected  number
of interactions  among partons  in a collision with impact parameter  $b$  when
the  colliding hadrons  are in configuration ${\mathbb{C}_1}$ and ${\mathbb{C}_2}$
(the  dependence on the c.m. energy has  been left implicit).
The difference with respect to the  discussion made  above  is that
in the previous case one was  discussing only a  subclass
of (hard)  interactions,  while  here  one refers  to the 
{\em total}  number of  parton interactions.

At this point we can    again make the factorization 
hypothesis, that the  fluctuations   in the  number 
of interactions    at different  impact parameters,
being related  to  the   distribution of  parton configurations
is  independent from $b$, and therefore 
one has:
\begin{equation}
\int d{\mathbb{C}_1}  ~\int d{\mathbb{C}_2} ~ 
P_{h_1} ({\mathbb{C}_1})
~P_{h_2} ({\mathbb{C}_2})~
\exp \left [ -\frac{ n(b, {\mathbb{C}_1},{\mathbb{C}_2})}{2} \right ]
= \int_0^\infty d\alpha~p(\alpha)
\exp \left [ -\frac{ \langle n(b)\rangle \, \alpha }{2} \right ]
\label{eq:ident2}
\end{equation}
As before the function $p(\alpha)$ satisfies the two  integral relations:
$$
\int_0^\infty d\alpha ~p(\alpha) = 1
~~, ~~~~~~
\int_0^\infty d\alpha ~\alpha~p(\alpha) = 1 ~.
$$

Using the identities  (\ref{eq:ident1})  and (\ref{eq:ident2})
we  can  now  rewrite in  manageable  form  equations 
(\ref{eq:siga_abs},\ref{eq:siga_el},\ref{eq:siga_difft},\ref{eq:siga_tot})
as:
\begin{equation}
\frac{d^2 \sigma_{\rm abs}}{d^2 b} = 1 - \int_0^\infty d\alpha ~p(\alpha) ~
e^{-\langle n(b,s)\rangle \, \alpha}
\label{eq:sig_abs}
\end{equation}
\begin{equation}
\frac{d^2 \sigma_{\rm el}}{d^2 b} =  \left [
\int_0^\infty d\alpha ~p(\alpha) ~
\left (1 - e^{-\frac{\langle n(b,s)\rangle \, \alpha}{2}} 
\right )
\right ]^2
\label{eq:sig_el}
\end{equation}
\begin{equation}
\frac{d^2 \sigma_{\rm diff+el}}{d^2 b} =  \int_0^\infty d\alpha ~p(\alpha) ~
\left [1 - e^{-\frac{\langle n(b,s)\rangle \, \alpha}{2}} \right ]^2
\label{eq:sig_Diff}
\end{equation}
\begin{equation}
\frac{d^2 \sigma_{\rm tot}}{d^2 b} =  2\;
\int_0^\infty d\alpha ~p(\alpha) ~
\left (1 - e^{-\frac{\langle n(b,s)\rangle \, \alpha}{2}} 
\right )
\label{eq:sig_tot}
\end{equation}

The   elastic scattering  amplitude is given by:
\begin{equation}
F_{\rm el} (q,s) 
 = i ~\int \frac{d^2b}{2 \, \pi}  ~ e^{i \vec{q}.\vec{b}}~
\int_0^\infty d\alpha ~p(\alpha) ~ \left [ 1 - e^{-\frac{\langle n(b,s)\rangle \; \alpha}{2}}  \right ]
\label{eq:dsigdt}
\end{equation}

It  is  important to  note  that
in  the limit  $p(\alpha) \to \delta [\alpha -1]$,
that is in the limit   where  one  neglects  the effects  of different 
parton configurations,  one has that  equations
  (\ref{eq:sig_el})  
and 
  (\ref{eq:sig_Diff}), that describe the elastic, and elastic + diffractive
cross sections  become   identical, that is
inelastic  diffraction  vanishes.  
Moreover, in this case,    the  expressions  for the absorption, elastic and total  
cross section  coincide with the
expressions  of the  simple eikonal  model  that  neglects  inelastic  diffraction.

%The  inelastic diffractive processes can be  studied  experimentally
%considering the  3--times differential  cross section:
%$d^3 \sigma_{\rm diff}/{(dM_1~dM_2 ~dt)}$  where  $M_1$  and $M_2$ are
%the masses of  the excited  states in  the    forward  and backward   regions and $t$ is 
%the 4--momentum  transfer.
%Inelastic projectile diffraction is obtained  fixing the target  mass to 
%$M_2 =  m_{\rm target}$   and  integrating  over 
%all $M_1$  larger  than  the projectile  mass (and similarly for target diffraction).
%Double  diffraction is  obtained    summing  over  all masses
%$M_1$ and  $M_2$  different  from the masses of the initial  state particles.

Equation (\ref{eq:sig_Diff}) sums over  all  diffractive  channels, and therefore
loses  all information about the  distributions of the excited  masses.
It  remains   however possible  to compute the  $t$  distribution
for elastic  scattering  and  for 
 inelastic  diffractive  scattering  (summing  over all possible open channels):
\begin{equation}
\frac{d\sigma_{{\rm  diff}+{\rm el}} } {dt} 
= \sum_{m}
\frac{d\sigma_m} {dt} =
\int dM_1 ~\int dM_2 ~
\frac{d^3\sigma_{\rm  diff}} {dM_1 \, dM_2 \, dt} 
\end{equation}

The differential   cross section for elastic  scattering
 $d\sigma_{\rm el}/dt$  can be  calculated 
from equations (\ref{eq:gen_dsig}) and 
(\ref{eq:dsigdt}):
\begin{equation}
\frac{d\sigma_{\rm el}}{dt} = 
\pi \;
\left [ \int_0^\infty  db~b ~J_0 (b \, \sqrt{|t|}) ~
\int_0^\infty d\alpha ~p(\alpha) ~
\left (1 -e^{-\frac{\langle n(b,s) \rangle \; \alpha}{2}} \right ) 
\right ]^2
\label{eq:diffr1}
\end{equation}
Similarly, the differential   cross section for  diffraction plus  elastic
scattering   can be calculated
as:
\begin{equation}
\frac{d\sigma_{{\rm diff}+{\rm el}}}{dt} = \pi \;
\int_0^\infty d\alpha ~p(\alpha) ~
\left [
 \int_0^\infty db~b ~J_0 (b \, \sqrt{|t|}) ~
\left (1 -e^{-\frac{\langle n(b,s) \rangle \; \alpha}{2}} \right ) 
\right ]^2
\label{eq:diffr2}
\end{equation}

It is  straightforward to see  that for $p(\alpha) = \delta [\alpha -1]$ 
expressions  (\ref{eq:diffr1}) and (\ref{eq:diffr2})  become  identical
and  equal to  the well known expression for  the simple  eikonal model:
\begin{equation}
\left . \frac{d\sigma_{\rm el}}{dt}  \right |_{\rm simple} = 
\pi \;
\left [
 \int db~b~ ~J_0 (b \, \sqrt{|t|}) ~
\left (1 -e^{-\frac{\langle n(b,s) \rangle}{2}} \right ) 
\right ]^2
\label{eq:diffr3}
\end{equation}

The  slopes  at $|t| =0$    ($B_{\rm  el}$ and  $B_{\rm diff}$)
can  be  calculated   from  the  definition (\ref{eq:Bdef}) 
(and the analogous  for   inelastic diffraction).
For  example,  using:
\begin{equation}
\lim_{t \to 0} \frac{d}{dt } \left [ J_0 (b \, \sqrt{|t|} ) \right ] =
-\frac{b^2}{4}
\end{equation}
one  obtains for $B_{\rm el}$:
\begin{equation}
B_{\rm el} = 
\left [\int_0^\infty  db ~ \frac{b^3}{2} ~
\int_0^\infty d\alpha ~p(\alpha) ~
\left (1 -e^{-\frac{\langle n(b,s) \rangle \; \alpha}{2}} \right ) 
\right ]
\times
\left [\int_0^\infty  db ~ b ~
\int_0^\infty d\alpha ~p(\alpha) ~
\left (1 -e^{-\frac{\langle n(b,s) \rangle \; \alpha}{2}} \right ) 
\right ]^{-1}
\label{eq:bel1}
\end{equation}
A  similar  expression for $B_{\rm diff}$ is  easily  derived.

\subsection{Comparison  with multi--channel  eikonal  formalism}
\label{sec:comp0}

A  method  to  apply to Good Walker ansatz    is  to construct explicitly a
a transition  operator as  a  matrix of $n \times n$--dimensions.
In this way   the   formal  indices  $m$  and $j$   in  section~\ref{sec:good}
become  simply integer indices  running from 1 to  $n$. 
To show the  mathematical equivalence between  this  multi--channel  
method and the  use of the effective configuration probability distribution
$p(\alpha)$ is   straightforward.
In  the multichannel approach  the  profile  function  $\Gamma_{\rm el} (b,s)$ 
is replaced  by  the ($n \times n$) matrix  $\hat{\Gamma}(b,s)$ that can be expressed in terms
of  the  eikonal  matrix  $\hat{\chi}(b,s)$:
\begin{equation}
\hat{\Gamma} (b,s) = 1 - \exp [-\hat{\chi} (b,s) ]
\end{equation} 
 The eigenvalues of $\hat{\Gamma}$  ($\Gamma_j$) and  $\hat{\chi}$ 
($\chi_j$)  can be  written  in the  form: 
\begin{equation}
\Gamma_j = 1 -e^{-\chi_j} = 1 - \exp \left [ - \frac{\langle n(b,s) \rangle}{2}
 \; \alpha_j \right ]
\end{equation}
The  interpretation of $\Gamma_j$ as   the effect of  absorption 
suggests  that $\langle  n(b,s)\rangle$ and all $\alpha_j$'s  are
real  and positive. 
In the  optical  analogy one  interprets the quantity:
$$
1 - [1 - \Gamma_j]^2 = 1 - e^{-2 \, \chi_j}  = 1 - e^{-\langle n (b,s) \rangle \; \alpha_j} 
$$
as the absorption probability  for the  eigenstate   $|\psi_j\rangle$ 
that corresponds to eigenvalue $\alpha_j$.
Assuming a 
Poisson probability distribution, the 
quantity $[\langle n(b, s) \rangle \; \alpha_j]$  can then  be interpreted  as
the average number of elementary interactions   for the eigenstate $|\psi_j\rangle$.
Without loss of generality, reabsorbing a constant in
the definition of  $\langle  n(b,s)\rangle$,  one can  impose the constraint
\begin{equation}
\sum_j p_j \; \alpha_j = 1
\label{eq:pnorm1}
\end{equation} 
where the quantities $p_j$'s:
\begin{equation}
p_j = \left | \langle  \psi_j | \varphi_{\rm initial} \rangle \right |^2
\end{equation}
measure the probability  overlaps   between 
the  initial  state  $|\varphi_{\rm initial}\rangle$  and the  eigenstates
$|\psi_j\rangle$   of the  transition matrix.

The  normalization  condition  (\ref{eq:pnorm1})  allows
to  interpret   the  quantity $\langle  n(b,s) \rangle$ as the
average  number   of elementary interactions 
(at impact parameter $b$ and  c.m. energy $\sqrt{s}$)  for the 
initial  state  $|\varphi_{\rm  initial}\rangle$.
It is then  possible to  define  the function $p(\alpha)$ as
\begin{equation}
p(\alpha) = \sum_j p_j ~\delta [\alpha - \alpha_j]
\label{eq:palphadef1}
\end{equation}
and to use  this  function 
to express  the results for the   total, elastic, absorption and diffractive
cross sections  as  integrals over  $\alpha$  according to  equations
(\ref{eq:sig_abs}--\ref{eq:sig_tot}).

It can be instructive to consider  an  explicit  model
that  implements the multi--channel model.
The minimum model that  includes  the 4~distinct processes  
of elastic scattering  together with 
target, projectile and double  diffraction,  must obviously 
consider a 4--dimensional Hilbert space  spanned  by 
4~physical eigenstates $|\varphi_{m}\rangle$
(that without loss of generality can symbolically be labeled 
as:  
$|\pi p\rangle$,
$|\pi \Delta\rangle$,
$|\rho p\rangle$ and
$|\rho \Delta\rangle$).

The  structure of this most general Good--Walker model
with 4--channels has been presented in \cite{Ahn:2009wx} 
and  is  discussed in detail  in appendix~\ref{sec:4chan}.
The most general  4--channel eikonal  
(for  a fixed  $\sqrt{s}$) is 
described  by  the  impact parameter
multiplicity distribution  $\langle n(b,s)\rangle$  and by 
the  $4 \times 4$   matrix $ \hat{M}$  that 
in the most general  case (where the  two initial state 
hadrons  are not identical  as in $\pi p$ scattering)
is  defined  by   4~real parameters  (in the case
where the  two interacting  hadrons
are identical,  as in $pp$ scattering, the parameters reduce to 2).

The eigenvalues and eigenvectors of the most general  matrix 
$\hat{M}$  are easily calculated  (as  show in   appendix~\ref{sec:4chan}).
Having solved  this  diagonalization problem, 
it is simple to obtain  the cross  sections  for 
the total, elastic, absorption and  
diffractive scattering.
The cross sections  depend on $\langle n(b,s)\rangle$ and
on the  4 (or 2 for  identical  initial particles)  parameters
of the matrix  $\hat{M}$.
The cross sections  can be recast in the form
(\ref{eq:sig_abs}--\ref{eq:sig_tot})  as  integrals over
$\alpha$ defining  $p(\alpha)$  according to equation (\ref{eq:palphadef1}) 
(where the  summations   over  $j$ now  runs   from  1 to 4).

Note that  in a   $n$--channel   eikonal,
the inelastic diffractive cross sections  is  obtained  as
an explicit sum of $n-1$ terms:
\begin{equation}
\sigma_{\rm diff} = \sum_{m \ne 1}  ~\sigma_m
\end{equation}
(we  are identifying  the state  $m=1$ with the  initial  state).
In  a  realistic discussion the  index  $m$  should  run  continuously  
over all possible diffractively excited states.
For the  4--channel  model
the  3 states   can  be  identified as representing
target, projectile and  double  diffraction.

It is  interesting to study, in the  framework of this  most general
4--channel model,  the relative importance of
single  versus  double  diffraction.
Considering for  simplicity the  case of   $pp$ collision,
the 4--channel Good--Walker  model has  two  free  parameters
($\beta_p$ and  $\epsilon_p$).
The relative  importance of  the elastic  and  diffractive  processes
can   vary  significantly    with  variation  of the model  parameters,
and  similarly  the   ratio  between  the
double  and  single  diffraction cross section   can  assume  different  values,
however   numerical  studies  show  that to a good  approximation 
(for the scattering of identical  particles)  one has:
\begin{equation}
\frac{\sigma_{\rm TD}}{\sigma_{\rm el}} 
\equiv \frac{\sigma_{\rm BD}}{\sigma_{\rm el}} 
\simeq \frac{\sigma_{\rm DD}}{\sigma_{\rm TD}} 
\equiv \frac{\sigma_{\rm DD}}{\sigma_{\rm BD}} 
\label{eq:ddappr}
\end{equation}
that is the  ratios of   cross sections 
 single (beam or target) diffraction   to 
the elastic one  is approximately equal to the ratio
of the double  diffraction  cross section to the single  diffraction one.
This  result  can also be  understood from inspection
of the  structure of the matrix  $\hat{M}$.
For negliglible $\epsilon_p$ for example, the relative
importance of the cross sections for elastic, single diffractions
and double diffraction scattering are in the ratio
$1:\beta_p:\beta_p^2$.
This   result  is  compatible  with the available data 
on double  diffraction (taking into account the large errors).

The separate calculation of the different components 
of the diffractive cross sections is a significant
merit for the 4--channel  model.
 A  limit of such an approach  is 
that it predicts  the multiplicity  distribution
of the elementary interactions
as the  superposition of 3 (for the scattering of identical  particles)
or 4 (in the general case) Poissonian  distributions of  different
average  values,  such  distribution might  be not sufficiently smooth
for a  realistic  comparison  with data.
The approach   of using  the  function 
of  a  real positive variable $p(\alpha)$
allows  to  consider  implicitly an  infinity of inelastic  channels. 

In the next  section we  will propose a  simple
parametrization of $p(\alpha)$  that  depends on a single parameter.

\section{Explicit Model}
\label{sec:explicit}
Equations 
  (\ref{eq:sig_abs}--\ref{eq:sig_Diff})   
allow to  compute the  total, elastic, diffractive and  absorption  cross 
sections in  terms of the 
impact parameter  multiplicity  distribution
 $\langle n (b,s) \rangle$  
(the average  number of elementary interactions at impact parameter 
$b$ and c.m. energy $\sqrt{s}$) and of the function $p(\alpha)$.
Unfortunately  the  shape of the function $p(\alpha)$ is  not  determined,
all  we have been able to establish is  that the 
first two moments  $\langle \alpha^k\rangle$ 
with $k = 0,1$ must be  unity.
It is  however reasonable to expect that the   most important 
property of the   function $p(\alpha)$ is its
second  moment  $\langle \alpha^2 \rangle$  or equivalently 
its  width $\sigma_\alpha^2 = \langle \alpha^2 \rangle -1$.

\begin{figure}[hbt]
\begin{center}
 \includegraphics[width=14.cm]{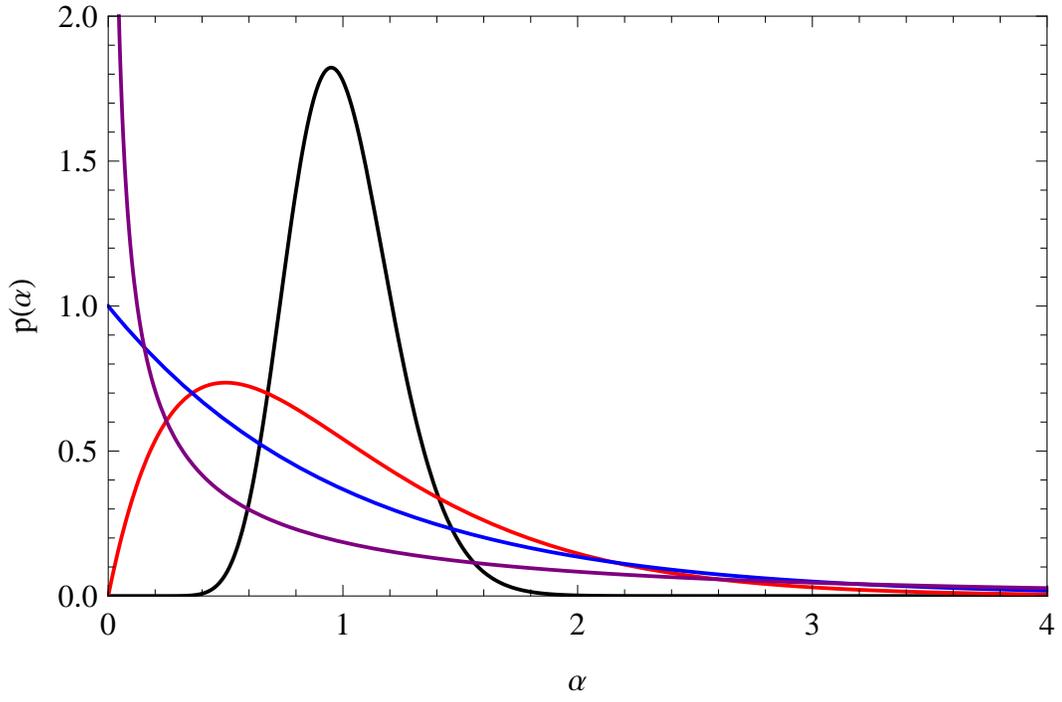}
\end{center}
   \caption{\small  
Plot of the function  $p(\alpha)$  
given in   equation (\protect\ref{eq:p_model}) 
for  four  values of the  parameter  $w$  ($w = 0.05$, 0.5, 1, 3).
For $w \to 0$  the function   takes the form  $\delta [\alpha -1]$.
\label{fig:palpha}  }
\end{figure}

Given  this  lack of  knowledge  about the shape of
$p(\alpha)$ we have chosen for it  a
simple analytic expression, that  allows
easy manipulations:
\begin{equation}
p(\alpha) = 
\frac{1}{w \; \Gamma \left (\frac{1}{w} \right ) } 
~ \left ( \frac{\alpha}{w} \right )^{\frac{1}{w}-1}
 \exp \left [-\frac{\alpha }{w} \right ] 
\label{eq:p_model}
\end{equation}
The $k$--th moment of the distribution is:
\begin{equation}
\langle \alpha^k \rangle  = \int_0^\infty d\alpha ~\alpha^k 
~p(\alpha ) = 
\frac{w^k \Gamma \left(k+\frac{1}{w}\right)}{\Gamma \
\left(\frac{1}{w}\right)}
\end{equation}
therefore one  finds:
\begin{equation}
\langle \alpha^0 \rangle = 1, 
~~~~~~
\langle \alpha \rangle = 1, 
~~~~~~
\langle \alpha^2 \rangle = 1 + w, 
~~~~~~
\sigma_\alpha^2 = w  ~~.
\label{eq:alphamom}
\end{equation}
Therefore the  parameter $w$ describes the   variance of the $\alpha$ 
distribution.
The integer  values   moments  for $n \ge 3$  can be written  also as:
\begin{equation}
%\langle \alpha^3 \rangle = (1+w) \, (1 + 2 \, w), ~~~~~
\langle \alpha^n \rangle = (1+w)  \ldots (1 + (n-1) \, w)
\end{equation}
Numerical  examples  of the  function $p(\alpha)$ are shown in figure~\ref{fig:palpha}.

An attractive  property of the  functional  form (\ref{eq:p_model})
 for  $p(\alpha)$ is  that it  allows to perform  analytically 
the integrations over $\alpha$  
in equations   (\ref{eq:sig_abs}--\ref{eq:sig_tot})   
to obtain  the quantities of  interest.
The profile  function  $\Gamma_{\rm el} (b,s)$  becomes: 
\begin{equation}
 \Gamma_{\rm el} (b,s) = 
\int_0^\infty d\alpha ~p(\alpha) ~ \left [ 1 - e^{-\frac{\langle n(b,s)\rangle \; \alpha}{2}}  \right ] =
1 -  \left (1 +  \frac{\langle n(b,s) \rangle \, w}{2} \right )^{-\frac{1}{w}}
\label{eq:wprof}
\end{equation}
The   expressions for the total, elastic  and diffractive
cross sections 
% (that is equations 
%  (\ref{eq:sig_tot}), 
%  (\ref{eq:sig_el})  and   
%  (\ref{eq:sig_Diff})) 
become:
\begin{equation}
\frac{d^2\sigma_{\rm tot}}{d^2b} = 
2 - 2 \left (1 +  \frac{\langle n(b,s) \rangle \, w}{2} \right )^{-\frac{1}{w}}
\label{eq:wtot}
\end{equation}
\begin{equation}
\frac{d^2\sigma_{\rm el}}{d^2b} = 
\left (1 -  \left (1 + \frac{\langle n(b,s) \rangle \, w}{2} \right )^{-\frac{1}{w}}  \right )^2
\label{eq:wel}
\end{equation}
%\begin{equation}
%\frac{d^2\sigma_{\rm diff+el}}{d^2b} = 
%1 - 2 \; \left (1  + \frac{\langle n(b,s) \rangle \, w}{2} \right )^{-\frac{1}{w}} +
%\left  (1 +  \langle n(b,s) \rangle \, w \right )^{-\frac{1}{w}} 
%\label{eq:wDiff}
%\end{equation}
\begin{equation}
\frac{d^2\sigma_{\rm diff}}{d^2b} = 
 \left (1  + \langle n(b,s) \rangle \, w \right )^{-\frac{1}{w}} -
\left  (1 +  \frac{\langle n(b,s) \rangle \, w}{2} \right )^{-\frac{2}{w}} 
\label{eq:wdiff}
\end{equation}

Using for    $\langle n(b,s)\rangle$  
the parametrization of equation (\ref{eq:pareik}) the expressions 
(\ref{eq:wtot}--\ref{eq:wdiff}) allow  to compute 
the different components of
hadron--hadron interactions  for  any given value
of $\sqrt{s}$ in terms  of three parameters: ($\sigma_{\rm eik}$,$r_0$,$w$).
It should be  noted  that the simple eikonal  model
corresponds to the  case $w\to 0$ and is  therefore included
as a limiting case of our model.

As a  critical  remark  we note  that  the qualitative  idea behind 
our one--parameter  modeling of $p(\alpha)$  is 
that   the most  important   feature of  $p(\alpha)$ is  its
second  moment $\langle  \alpha^2 \rangle = w + 1$. 
This  however  is only true in first approximation.
Functions   $p(\alpha)$   that  differ  only  for 
moments  $\langle \alpha^k \rangle$  with $k > 2$  can  also 
produce   different  cross sections.  An  example of  this behaviour
is the 4--channel  model of appendix~\ref{sec:4chan}.
For  $pp$ interactions this  model has 
two  free  parameters ($\beta_p$ and  $\epsilon_p$).
The variance of $p(\alpha)$ in the model   
is  $(1+\beta_p)^2 -1$ is   uniquely  determined  by  $\beta_p$,
however  the  value of the different  cross sectios
(elastic, diffractive and absorption)  depend on both   of the 
model  parameters.

The  single  parameter  description of $p(\alpha)$ of 
equation (\ref{eq:p_model})  seems  in any case a  reasonable form 
to  investigate phenomenologically the consequences 
of  equations (\ref{eq:sig_abs}--\ref{eq:sig_tot}).

%The  comparison of the data
%of  $\sigma_{\rm tot}$,   $\sigma_{\rm el}$ and  $\sigma_{\rm diff}$
%at a  certain   c.m.  energy  $\sqrt{s}$  with 
%predictions  based on  equations (\ref{eq:wtot}--\ref{eq:wdiff})
%allows  to   extract the values of the 3 parameters  ($\sigma_{\rm eik}$,$r_0$,$w$).
%Such a parameter  determination will be  performed in section~\ref{sec:parameters}.

\subsection{Parameter determination}
\label{sec:parameters}

Equations (\ref{eq:wtot}--\ref{eq:wdiff}) allow  to  determine
(at a  certain $\sqrt{s}$) 
the  set  ($\sigma_{\rm eik}, r_0,w$)  of  the three parameters  in   the model
from  measurements  of  $(\sigma_{\rm tot},B,\sigma_{\rm diff})$.
As an example of this parameter  determination 
we  discuss here in  some detail the
measurements performed  at one  particular  value of the c.m. energy 
($\sqrt{s} = 1.8$~TeV)  by one  detector
(CDF  at the Fermilab $\overline{p}p$ collider).
The CDF  experiment \cite{Abe:1993xy,Abe:1993xx} has measured
$\sigma_{\rm tot} \, (1+ \rho^2)  =  81.83 \pm 2.29$~mbarn
(that  estimating  $\rho \simeq 0.15$  corresponds to
$\sigma_{\rm tot} =  80.03 \pm 2.24$~mbarn);
an elastic  cross section 
$\sigma_{\rm el} = 19.7 \pm 0.85$~mbarn,
and a slope of the  forward  elastic cross section
$B_{\rm el}= 16.98  \pm 0.25$~GeV$^{-2}$.
In addition the CDF collaboration has measured \cite{Abe:1993wu}
the  single diffractive cross section: 
$\sigma_{\rm SD} = 9.46 \pm 0.44$~mbarn.
The three quantities  $\sigma_{\rm tot}$,   
$\sigma_{\rm el}$ and $B_{\rm el}$ are  related by
the  unitarity   relation (\ref{eq:unit1}), and therefore
only two of them  are independent.  In the  following
we will    fix our  attention on  $\sigma_{\rm tot}$ and  $B_{\rm el}$.

Our  formalism    allows  only the calculation of 
the  total  diffractive  cross section,  summing over 
single and  double  diffraction processes.  In order to compare
the model to the single diffraction measurement  of CDF 
we  have therefore to  include  some  estimate of double diffraction.
We will use  the result (\ref{eq:ddappr})  that  allows  to
estimate  the  complete  diffractive  cross section $\sigma_{\rm diff}$ from
the measurements of the elastic and single  diffractive  one  as:
\begin{equation}
\sigma_{\rm diff} = 
\sigma_{\rm SD} +
\sigma_{\rm DD}  \simeq \sigma_{\rm SD} 
~ \left ( 1 + \frac{\sigma_{\rm SD} }{2 \, \sigma_{\rm el}}  \right )
\end{equation}
This  hypothesis   leads  us  to estimate $\sigma_{\rm diff}$
at $\sqrt{s} = 1800$~GeV  from the CDF  data  as  approximately
11.6~mbarn.

Neglecting  the measurement of  the diffractive  cross section
and considering   only  the experimental results for $\sigma_{\rm tot}$ and
$B_{\rm el}$, there  is an  infinity  of sets  of  parameters
 ($\sigma_{\rm eik}, r_0,w$)   that reproduce the data.
This infinity of  solutions  can be parametrized  by the  value 
of  $w$, that  can take  any   non--negative  value.

The limiting case  $w=0$  corresponds to the 
simple eikonal  model.
For  $w=0$  the  values of the parameters
that  reproduce the central value of the CDF 
measurements  are:  $\sigma_{\rm eik} \simeq 124.1$~mbarn  
and $r_0 \simeq 0.2527$~fm   ($\simeq 1.08 ~R_p$).
To this ($w = 0$) solution corresponds  a 
vanishing  diffractive  cross section.

\begin{figure}[hbt]
\begin{center}
 \includegraphics[angle=90,width=12.5cm]{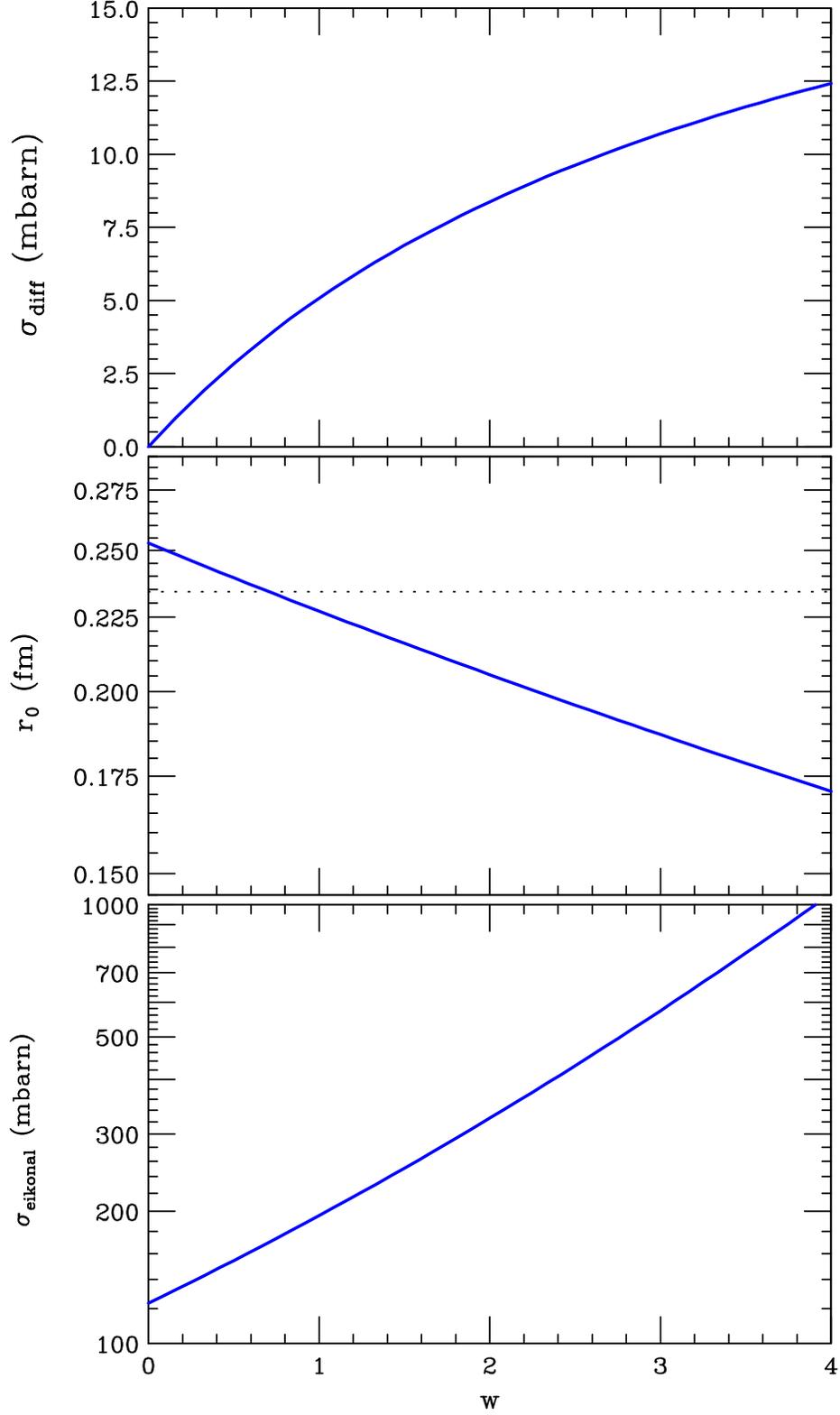}
\end{center}
   \caption{\small  
The  middle  and bottom  panel  show  the
triplet of  parameters ($w$, $\sigma_{\rm eik}(w)$ and $r_0(w)$) 
that reproduce 
(using equations  (\ref{eq:wtot}) and (\ref{eq:wel})   with
expression (\ref{eq:pareik})  for $\langle n(b,s)\rangle$)
the  measurements  of
$\sigma_{\rm tot}$ and $B_{\rm el}$  
obtained  by CDF  \protect\cite{Abe:1993xy,Abe:1993xx}
  at $\sqrt{s} = 1.8$~TeV
(note the logarithmic  scale   in the bottom panel for $\sigma_{\rm eik}$).
The   top panel  shows the corresponding    value of the   diffractive  cross section.
\label{fig:exa}  }
\end{figure}

The  solutions  $[w, \sigma_{\rm eik} (w),r_0(w)$]     that  reproduce
the   central value of  the CDF  measurements for
$\sigma_{\rm tot}$ and  $B_{\rm el}$ at  $\sqrt{s} = 1800$~GeV  are shown  in
fig.~\ref{fig:exa}. 
Increasing $w$   the value of  $\sigma_{\rm eik}(w)$  
grows monotonically, while $r_0(w)$    decreases. 
The triplets  \{$w, \sigma_{\rm eik} (w), r_0 (w)$\}  
result  in identical  $\sigma_{\rm tot}$ and $B_{\rm el}$, but  produce
different  diffractive  cross sections, with $\sigma_{\rm  diff} (w)$ 
growing  monotonically with $w$    as  also shown in 
fig.~\ref{fig:exa}.  
The  value $\sigma_{\rm diff} = 11.6$~mbarn is  obtained  for $w  = 3.48$.

Perhaps the  most  striking feature of  figure~\ref{fig:exa}  is the
rapid increase  of $\sigma_{\rm eik}$ with  $w$.
As an  illustration, 
for   $w \simeq  3$  (that results  in  $\sigma_{\rm  diff} \simeq 10.7$~mbarn)
one needs  a  smaller $r_0$   ($r_0 = 0.186$~fm)
and $\sigma_{\rm eik} \simeq 580$~mbarn, that  is almost 5 times  larger  
than the   value of  $\sigma_{\rm  eik}$  that reproduces the  measurements
for  $w =0$.  

\begin{figure}[hbt]
\begin{center}
 \includegraphics[angle=90,width=15cm]{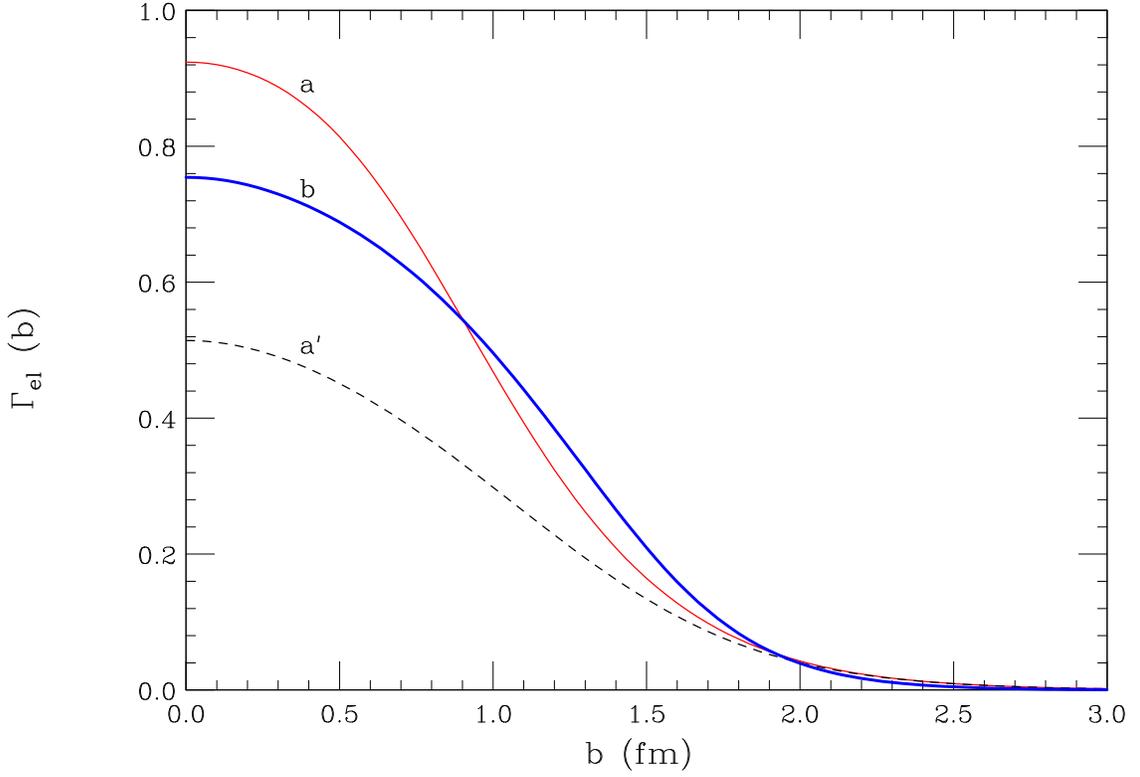}
\end{center}
   \caption{\small  
Profile function $\Gamma_{\rm el} (b)$ for 
$pp$ scattering  as a function of the impact parameter $b$.
The curve $(a)$  is  calculated   in the simple  eikonal
of  equation (\protect\ref{eq:ansatz1}),  using for the  $\langle n(b,s)\rangle$ the
parametrization (\protect\ref{eq:pareik})  with $\sigma_{\rm eik} = 124$~mbarn
and $r_0 = 0.253$~fm.  The corresponding values of 
$\sigma_{\rm tot}$ and  $B_{\rm el}$ are 
$\sigma_{\rm tot} = 80.3$~mbarn and $B_{\rm el}=16.98$~GeV$^{-2}$.
Curve  ($a^\prime$)  is  calculated  assuming the same  interaction profile 
$\langle  n(b,s) \rangle$  (that is the same  parameters $\sigma_{\rm eik}$ and
$r_0$)   as   for  curve ($a$) but using the model of  equation (\ref{eq:wprof})
for the profile   with the   form (\ref{eq:p_model}) for $p(\alpha)$ with 
$w = 3$. The  resulting  profile function  is  
smaller (implying a smaller $\sigma_{\rm tot}$)  and broader
(implying a larger  $B_{\rm el}$).
The profile $(b)$  is calculated  using  the same  model used for   curve ($a^\prime$)
with the same  value  $w = 3$, however  the parameters
that  describe  $\langle n(b,s) \rangle$ are now
$\sigma_{\rm eik} = 582$~mbarn  and $r_0 =  0.186$~fm.
The  profile function  ($b$)  results  in the same
$\sigma_{\rm tot}$ and $B_{\rm el}$ as  curve ($a$). 
\label{fig:profile}  }
\end{figure}

For  a qualitative understanding of these results it can be instructive
to consider  figure~\ref{fig:profile}. 
Curve $(a)$  shows the profile function $\Gamma_{\rm el} (b)$  
that  corresponds to the  solution  with $w=0$  that we have just discussed.
Curve  ($a^\prime$) shows the  profile that is obtained  for 
the same  parameters  $(\sigma_{\rm eik}, r_0$),
that  is for the same  $\langle n (b,s) \rangle$, 
 of the $w=0$ solution, but  using the value  $w=3$.
The resulting profile  function
is  smaller (that is  produces a smaller $\sigma_{\rm tot}$)  and broader
(implying a larger  $B_{\rm el}$).  These  features  can  be  readily understood
from inspection of  equation (\ref{eq:wprof}).
In order to  obtain the desired  values of $\sigma_{\rm tot}$ and $B_{\rm el}$ 
using  the model (\ref{eq:wprof})  and $w = 3$ one  needs to 
modify the  quantity  $\langle n(b,s) \rangle$, choosing both a 
larger  $\sigma_{\rm eik}$  to  increase the area  under the profile,
and  a smaller $r_0$ to obtain the desired  value of $\langle b^2\rangle \propto B$.
The  solution  is  shown  in fig.~\ref{fig:profile} as  curve ($b$).
The profile  functions of  curves ($a$) and ($b$)
in fig.~\ref{fig:profile}  produce identical    $\sigma_{\rm tot}$, and
identical $d\sigma_{\rm el}/dt$ for small $|t|$,  but differ in the
description of  elastic scattering at large $|t|$.

To summarize this  discussion: 
in  the simple  eikonal  model   the impact parameter  multiplicity distributions
$\langle n(b,s) \rangle$  
uniquely  defines  the profile function.
In our model the profile  function is  determined
(see  equation (\ref{eq:wprof}))  also by the function  $p(\alpha)$,  and
different  choices  for the  shape of  $p(\alpha)$  
produce   different   profiles, and  therefore 
different values of  the total and elastic  cross sections.
Viceversa, the estimate of  $\langle n(b,s)\rangle$
(or in terms  of   the parametrization (\ref{eq:pareik})  the  values
of $\sigma_{\rm eik}$   and $r_0$) 
that reproduces the measured  values
of  $\sigma_{\rm tot}$ and $\sigma_{\rm  el}$  (or  $\sigma_{\rm tot}$ and $B_{\rm el}$)
strongly depends on the assumptions made for $p(\alpha)$.

The  function $p(\alpha)$, and in particular
its width,  controls the size of the  inelastic diffractive  cross section,
therefore one can obtain  information about its  properties from
the  experimental  data on the rate of diffractive   events.
The  bottom line  is that it is essential  to  include in a consistent
way inelastic  diffraction  in the theoretical  framework
that  describes hadronic cross sections.

These considerations are the  main qualitative results of this  work:
the  consistent  introduction of  inelastic  diffraction
in the eikonal  formalism  results in:
\begin{enumerate}
\item  an eikonal  cross section $\sigma_{\rm eik}$ that
is  several times larger  than  estimates   based on the simple`
eikonal model that does not 
consider explicitly diffraction;
\item  a narrower distribution of hadronic  matter.
In the case of  protons,  this  distribution is estimated as 
narrower  than  the charge distribution  infered  
by the  electromagnetic  form factor.
\end{enumerate}

These effects  can   have  important  consequences 
in the prediction of  the properties
of   particle  production in high energy hadron  interactions, if
one takes into account the interpretation of the eikonal  
as  a description of   the multiple interaction structure of
the collision.  
In this case the ratio $\sigma_{\rm eik}/\sigma_{\rm inel}$ has the
physical meaning of the average number of elementary interactions
per inelastic  event, therefore   our results  imply 
that this average number of   elementary interactions 
is  several  times  larger  than previous  estimates.
The precise way to  relate this  quantity 
(the average number of elementary interactions per collision)  to observable  
quantities, such as  the multiplicity distribution,
depends  on a number of  additional  assumptions that  have to be  made
in a Montecarlo  modeling of multiparticle production.

The  theoretical  framework we are  considering predicts not only the
average number of  interactions in a collision,  but also  the 
detailed  multiplicity distribution for  such interactions in one  collision.

In section~\ref{sec:multiple} we have  given  in equation (\ref{eq:partial2})
the  multiplicity  distribution  of    the number of hard observable jets 
per collision, in terms  of  the   quantities
$\langle n_{\rm jet}(b,s) \rangle$
and  $p(\alpha)$.  
The  generalization  to the  multiplicity 
distribution of the total  number of  elementary interaction in a collision
can  be immediately obtained  replacing  
the average number of  hard interactions  at a fixed  impact parameter
and c.m. energy 
$\langle n_{\rm jet}(b,s) \rangle$
with   the average for the total number of
elementary interactions $\langle n(b,s) \rangle$.  
In addition, the most  economical  assumption
is to assume  that the  functions  $p(\alpha)$   relevant 
in the two cases are  (at least approximately)  equal.

For the simple functional  form  of $p(\alpha)$   given in
equation (\ref{eq:p_model})  the integrals over $\alpha$ in   
the analogous of equation (\ref{eq:partial2})
can  be performed, with the result:
\begin{equation}
\sigma_k = \frac{w^k}{k!} ~\Gamma \left (k +\frac{1}{w} \right ) 
~\left [\Gamma \left (  \frac{1}{w}\right ) \right ]^{-1} \int d^2 b~
\langle n(b,s) \rangle^k ~ \left \{ 1+ w \; \langle n(b,s)\rangle 
\right\}^{-\left ( k + \frac{1}{w} \right )} 
\end{equation}
This distribution  of the number of elementary interactions
should be inserted  in Montecarlo implementations
to have  predictions  for the charged  particles multiplicity 
and other  observable  quantities.

\section{Energy Dependence}
\label{sec:energy}
 The model  we have outlined  in the previous  sections  
consider the relation  between directly  observable  quantities
such as the  total and elastic cross sections and 
on the other hand the  eikonal  cross  section $\sigma_{\rm eik}$ and
the distribution of   hadronic  matter in the colliding  particles
(simply parametrized by the quantity $r_0$),   with an   additional  parameter
$w$ that  is related to the width of  the fluctuations  in the
parton configurations of the colliding  hadrons.
Taking into account these fluctuations  
allows  a consistent treatment of 
inelastic  diffraction. The  values of  $\sigma_{\rm eik}$
and $r_0$  that  correctly describe the  data, have a strong
dependence on the parameter $w$, and therefore on  the measured  values
of the  cross section for  inelastic diffraction.

\begin{figure}[hbt]
\begin{center}
 \includegraphics[angle=90,width=15cm]{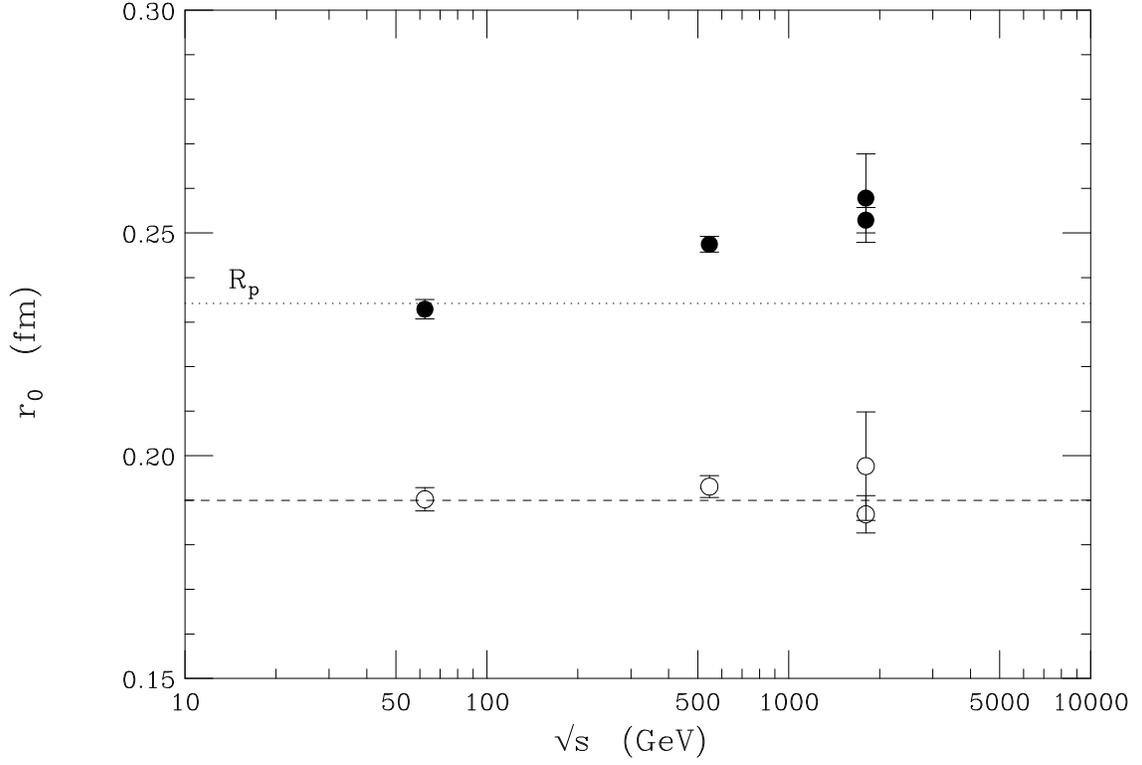}
\end{center}
   \caption{\small  
Values of the $r_0$ parameter  that
reproduce the experimental data  for  $\sigma_{\rm tot}$  and
$B_{\rm el}$  obtained  at the  ISR $pp$ collider
($\sqrt{s} = 62.3$~GeV), and at the Tevatron  $p\overline{p}$ collider
($\sqrt{s} = 546$~GeV  by CDF,   and  
$\sqrt{s} = 1800$~GeV  by CDF and E710).
The solid (empty)  points  are
calculated  for $w=0$ ($w=3$).
The dotted  line at $r_0 = 0.234$~fm  corresponds to the proton charge radius $R_p$.
The dashed line  corresponds to  the constant value $r_0 = 0.19$~fm, and is a reasonable
representation of the results  for $w=3$.
The  corresponding  values of $\sigma_{\rm eik}$ are shown in
fig.~\protect\ref{fig:fit_seik}.
\label{fig:fit_r0}  }
\end{figure}

\begin{figure}[hbt]
\begin{center}
 \includegraphics[angle=90,width=15.cm]{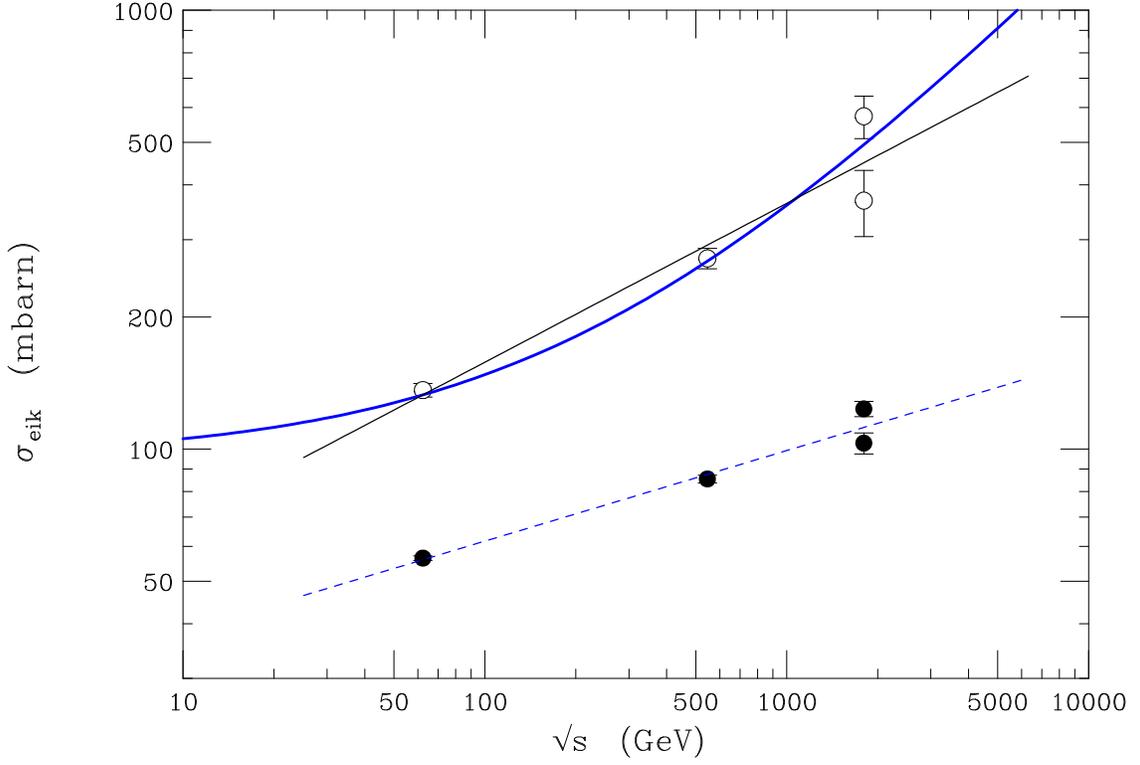}
\end{center}
   \caption{\small  
Values of the $\sigma_{\rm eik}$ parameter  that
reproduce the experimental data  for  $\sigma_{\rm tot}$  and
$B_{\rm el}$  obtained  at the  ISR $pp$ collider
($\sqrt{s} = 62.3$~GeV), and at the Tevatron  $p\overline{p}$ collider
($\sqrt{s} = 546$~GeV  by CDF,   and  
$\sqrt{s} = 1800$~GeV  by CDF and E710).
The solid (empty)  points  are
calculated  for $w=0$ ($w=3$).
The  corresponding  values of $r_0$ are shown in
fig.~\protect\ref{fig:fit_r0}.
The dashed  line is a power law  fit  
($\sigma_{\rm eik} (s) = K \; s^\alpha$)
to the results  for $w=0$.
The thin (black) line is a fit   
to the results for  $w=3$ with the same  power law  form.
The thick (blue) line is a fit  
to the same points with the form  $\sigma_{\rm eik} (s) = \sigma_0 + K \;s^{\alpha}$
with $\alpha = 0.35$.
\label{fig:fit_seik}  }
\end{figure}

A  calculation  of the evolution with energy of the 
hadronic cross section requires additional theoretical 
assumptions to predict the energy dependence of the 
model parameters.
To gain insight on this  problem, we have taken  a phenomenological
approach  and we have  considered  a  representative subset of the
available  high energy data.
A few high energy experiments  have  measured   both the total
and  elastic  cross section,  together  with  the forward slope $B_{\rm el}$.
These  results can  be  described
in terms of our 3--parameter $\{\sigma_{\rm eik}$, $r_0$, $w\}$ model,
using the same  approach discussed  for the CDF data  at $\sqrt{s} = 1800$~GeV.
The results   are  shown
in fig.~\ref{fig:fit_r0}  and~\ref{fig:fit_seik}.
In these   figures  the points  give  the values of   $r_0$ and $\sigma_{\rm eik}$ 
that  reproduce  the  measurements   of the pairs $(\sigma_{\rm tot},B)$  
using two assumptions   for 
the third  parameter:  $w=0$ and $w=3$.
The errors on the  estimates  
reflect only the experimental statistical errors.
At $\sqrt{s} = 1800$~GeV, one  has   two  independent  measurements
of the  total  cross section by the CDF  \cite{Abe:1993xy} and
E710 \cite{Amos:1992jw}   experiments, both at the  Fermilab  $p\overline{p}$ collider.
The point at $\sqrt{s} = 546$~GeV  is  also from  CDF,  while the point
at $\sqrt{s} = 62.3$ was  obtained  for $pp$ scattering 
at the CERN ISR collider \cite{Amos:1985wx}. 

The calculation with   $w=0$  corresponds to the  simple  eikonal model, 
and  in the   framework to our  model 
leads to a vanishing  inelastic  diffraction cross sections.
For each  pair  of experimental results $(\sigma_{\rm tot},B)$ 
we have  performed a scan of  $w$  similar
to the one  that we have   shown  in detail for the CDF point
at $\sqrt{s} = 1800$~GeV  (see also  fig.~\ref{fig:exa}), calculating
the pair  ($r_0(w),\sigma_{\rm eik}(w)$)  that   reproduces the
experimental  results.  This also  imply  a  value
$\sigma_{\rm diff}(w)$   obtained from  equation (\ref{eq:wdiff}).

At $\sqrt{s} = 1800$~GeV both the CDF \cite{Abe:1993wu} and
E710 \cite{Amos:1992jw}  have  measured the   single  diffractive  cross sections.
Averaging  with  equal  weight (to take  into account
large systematic  uncertainties)   and using the ansatz 
(\ref{eq:ddappr})  to estimate double  diffraction
we have  estimated  $\sigma_{\rm diff} \simeq  10.7$~mbarn.
This   value of the diffractive  cross section is  reproduced
in our  model (at the corresponding  energy  $\sqrt{s}  = 1800$~GeV)
with $w \simeq 3$.  Including a  20\%  uncertainity on the estimate 
of  $\sigma_{\rm diff}$,  $w$  can be estimated   at this  energy
as $w = 3^{+1.2}_{-0.9}$.

\begin{figure}[hbt]
\begin{center}
 \includegraphics[angle=90,width=15cm]{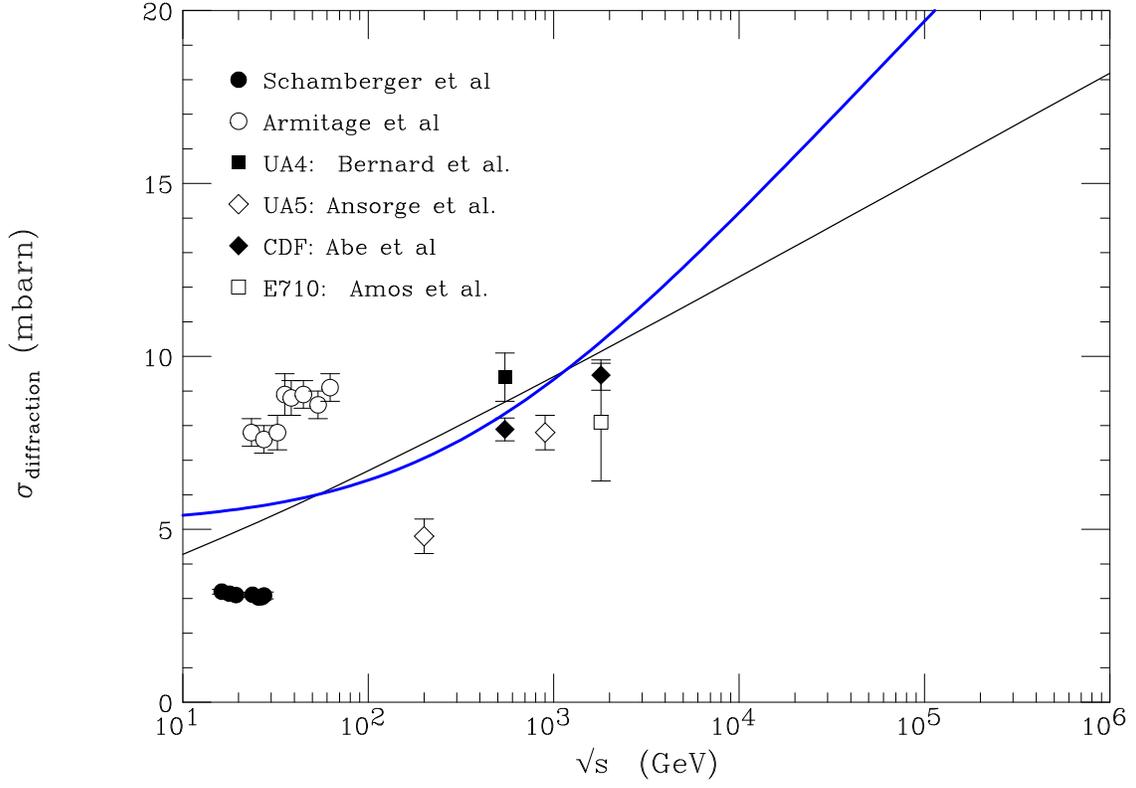}
\end{center}
   \caption{\small  Inelastic  diffraction cross
section calculated according  to equation (\protect\ref{eq:wdiff})
using constant  values  $w =3$ and $r_0 = 0.19$~fm. For the  thin (black) 
[thick (blue)]  curves 
we  have used for $\sigma_{\rm eik} (s)$ the fit  shown  with 
the corresponding lines in fig.~\protect\ref{fig:fit_seik}.
The experimental results  are for single  diffraction only
(Schamberger \cite{Schamberger:1975ea},
Armitage \protect\cite{Armitage:1981zp},
UA4 \protect\cite{Bernard:1986yh},
UA5 \protect\cite{Ansorge:1986xq},
CDF \protect\cite{Abe:1993wu},
E710 \protect\cite{Amos:1992jw}).
\label{fig:diffraction}  }
\end{figure}

The comparison  of the calculated diffractive cross section
with the data  is problematic because
the discrepancies between the different experimental 
results  (see fig.~\ref{fig:diffraction}) 
clearly indicate  the presence
of  significant  systematic  errors, moreover, 
as we have  discussed  before,
one  has the theoretical uncertainty  related  to the ratio
between the single and double diffraction contributions.

Our  numerical   studies indicate that  the choice of
an energy independent value  $w \simeq 3$,
that reproduces the  diffractive cross section
measured  at  the  Fermilab collider at $\sqrt{s} = 1800$~GeV, 
gives  in  fact a reasonably good description  of the 
experimental results on diffraction  at  all   energies.
The assumption of an energy independent  value of  $w$ is  
clearly the simplest one,  and  in  view of the fact that
it  produces a  reasonable agreement with the available  data
it will be made in the following.

It is interesting  to note  (see fig.~\ref{fig:fit_r0}), that 
assuming for  $w$  the  constant value  $w=3$
the  resulting values of the parameter $r_0$ 
are  also   consistent  with an energy independent   value
$r_0 \simeq  0.19$~fm.
In the simple  eikonal  model, as  discussed  before,
in order to  reconcile the growth of $\sigma_{\rm tot}$ and $B_{\rm el}$,
it is  necessary   to  increase the  width of the overlap  function 
 $A(b,s)$ with $s$, and therefore (using our  parametrization)
to increase $r_0(s)$.   The  necessity of this  growth is evident 
in  fig.~\ref{fig:fit_r0}.

The  model  we are discussing requires  
an overlap  function  that is  first of all
significantly  narrower than  previous estimates   
based on  the simple  eikonal model;
moreover  (and in contrast to the simple  eikonal model)
the  overlap function is  energy  independent.
It may  appear  surprising   that the overlap  function is
narrower than what is  estimated on the basis of 
the  proton charge  distribution.
A  possible  explanation is that the dominant  contribution to the 
overlap  function is the scattering between  soft gluons.
The  narrow  $A(b)$  predicted in this  model
therefore implies  that the  impact parameter distribution 
of soft gluons  is  (i)  narrower that the charge  distribution
(that is  presumably controled  by valence  quarks), 
(ii)  independent (for small $x$) from the $x$ of the gluons.
It should soon be  possible to  test  these hypothesis  with  studies
of the impact parameter  PDF's.

Figure~\ref{fig:fit_seik}  shows  the energy dependence of
the  third parameter  of our model, $\sigma_{\rm eik} (s)$. 
There are two  remarkable  features  in this   behaviour.
The  first  is  that (as  we have  already discussed in
section~\ref{sec:parameters})   the values  of $\sigma_{\rm  eik}$ needed
to describe the  experimental data  in a  model that  includes
diffraction  are significantly larger  
than estimates  based on the simple eikonal model, 
the second is  that  the 
growth of $\sigma_{\rm eik} (s)$ with  c.m. energy is  significantly
more  rapid.

The  two   straight lines in  fig.~\ref{fig:fit_seik} are power
law   fits  of   the estimated  values  of form $K \; s^\alpha$.
For the simple  eikonal model
($w=0$)  the power law  fit  
is $\sigma_{\rm eik} (s) = 23.8\; s^{0.10}$~mbarn
(with $s$  measured in GeV$^2$),
for the best fit  model  
($w=3$)  the power law  fit  
is $\sigma_{\rm eik} (s) = 29.7\; s^{0.18}$~mbarn.

The  choice of a  power  law  fit  is  however
clearly not  necessary, and  the extrapolation
of the  fit  at higher  (and lower) energy is therefore  very uncertain.
Motivated  by  fits of the $\sigma_{\rm  eik}(s) = \sigma_{\rm soft} + 
\sigma_{\rm hard} (s)$ 
that include an (approximately)   constant  
soft component  and
an energy varying  hard  component,  we  have  fitted  the data
with the form $\sigma_{\rm eik} (s)= \sigma_0 + K\; s^{0.35}$,   obtaining the result
$\sigma_{\rm  eik} (s)= 95 + 2.1 \; s^{0.35}$~mbarn (and  $s$ measured in GeV$^2$).

The  motivation for the  functional form  of this  fit
(that should  however also  be  considered  as  purely phenomenological)
is that 
integrating above an energy independent  $p_\perp^{\rm min}$, the  
jet cross  section     $\sigma_{\rm jet} ( p_\perp^{\rm min}, \sqrt{s})$ 
(see equation (\ref{eq:jet2a}))  has
qualitatively the  behaviour:
\begin{equation}
\sigma_{\rm jet} ( p_\perp^{\rm min}, \sqrt{s}) \propto
\frac{\alpha_s^2}{(p_\perp^{\rm min})^2}
~ \frac{\tau^{-\epsilon} \, (- \log \tau)}{\epsilon}
\propto 
\frac{\alpha_s^2}{(p_\perp^{\rm min})^2}
~ {s^{\epsilon} \,\log s}
\label{eq:asymp}
\end{equation}
where  $\tau = 4 \, (p_\perp^{\rm min})^2/s$, and
the  quantity $\epsilon$ is  related to the  behaviour of
the PDF's  for  $x\to 0$:
\begin{equation}
\lim_{x \to 0}  f(x)  \sim  \frac{1}{x^{1+\epsilon}}
\label{eq:asymp1}
\end{equation}
The behaviour 
(\ref{eq:asymp})  can be easily be  obtained
from the convolution of  PDF's  with the asymptotic form 
(\ref{eq:asymp1}).
Recent measurements  of the PDF's  at HERA 
\cite{Gluck:1994uf,Nadolsky:2008zw,Martin:1998sq,Martin:2009iq} have  shown  
that  their behaviour for $x\to 0$ can be  reasonably well
represented  with the functional  form (\ref{eq:asymp1})
and  $\epsilon \simeq  0.3$.

It is  interesting to note  that  it has  been  argued that the 
very fast  growth of the jet--cross section  with $\sqrt{s}$ implied  
by the  small  $x$  behaviour of the PDF's  is  problematic, and
in  fact unphysical.  
Note that in the simple eikonal  model,   an  energy dependence
of $\sigma_{\rm eik}(s)$  of type  $s^{0.3}$  (or faster)  is  not acceptable
because it   implies that $\sigma_{\rm tot} (s)$  grows  
with energy   more rapidly  than  the observations.
Following this  observation, the   rapid  growth  of  $\sigma_{\rm jet}(s)$ with
$s$  has  been  tamed assuming  that  the threshold  $p_\perp^{\rm min}$ of
applicability  of perturbation  theory  also grows  with $s$.  
This   growth  has  been connected to phenomena of 
``saturation'', or  screening  among the  partons.
In the  framework of  the model we are considering  a  fast 
growth of $\sigma_{\rm eik}(s)$ is  not only acceptable  but in fact 
necessary.  A simple  model where the growth of  $\sigma_{\rm eik}(s)$
is  explained with the dominant contribution of
a  minijet cross section, calculated  perturbatively
above an  energy independent $p_\perp^{\rm min}$  can provide
 a  $\sigma_{\rm eik}(s)$   with the  needed properties.
For  consistency, it is however  necessary 
that the effects of  saturation and parton  screening
are small.

The model  we are describing, already at $\sqrt{s} = 1800$~GeV  has
a ratio  $\sigma_{\rm eik}/\sigma_{\rm inel} \simeq 10$.  This   implies
that  the number of  elementary interactions in
an inelastic  collision at this energy is also  approximately 10.
This, at  first sight,  may appear too large. The  potential  danger 
is  that  this  large  average  number
of elementary interaction per collision
could  result in a  too large  average multiplicity 
and in  a too soft  inclusive  spectrum of particles in  the final state.
  These  questions can (and should) be 
addressed  properly   with a detailed  Montecarlo calculation, that
includes  a  modeling of  particle  production in the presence
of  different  numbers of  elementary interactions.

Figure~\ref{fig:diffraction}  shows  our  calculation of the diffractive  cross
section  $\sigma_{\rm diff}(s)$ including  the extrapolation to  high energy.
The calculation  is performed  with equation (\ref{eq:wdiff}),  using
energy  independent  values $w = 3$, $r_0 = 0.19$~fm  
and the two parametrization of $\sigma_{\rm eik}(s)$   shown  in fig.~\ref{fig:fit_seik}.

\begin{figure}[hbt]
\begin{center}
 \includegraphics[angle=90,width=15cm]{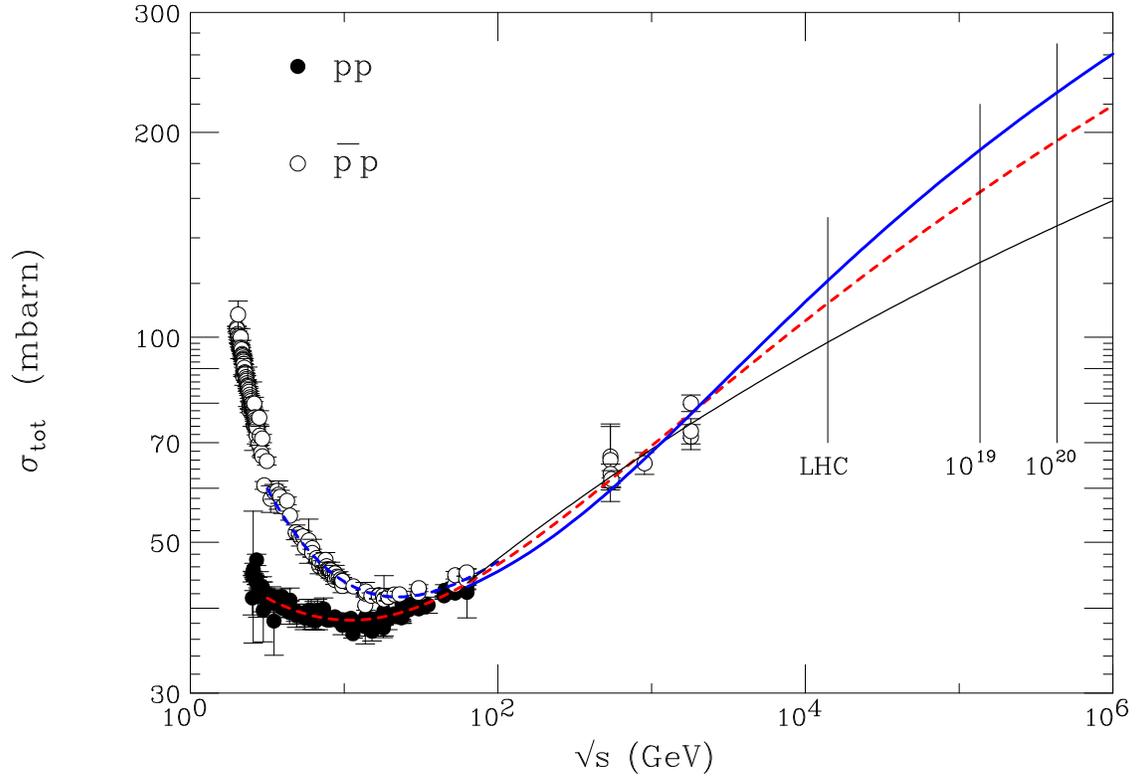}
\end{center}
   \caption{\small  The  points  are  measurements of the
$pp$ and $p\overline{p}$ total  cross sections.
The dashed lines  are  the fit   of $\sigma_{\rm tot} (s)$
suggested  in the  PDG \protect\cite{pdg}. 
The other  two lines are predictions
obtained  from equation (\protect\ref{eq:wtot})
using constant  values  $w =3$ and $r_0 = 0.19$~fm.
For the  thin (black) 
[thick (blue)]  curve 
we  have used for $\sigma_{\rm eik} (s)$ the fit  shown  with 
the corresponding lines in fig.~\protect\ref{fig:fit_seik}.
 \label{fig:stot}  }
\end{figure}

Figure~\ref{fig:stot}  shows the result of our model  for the  total  cross section,
comparing with the  available data  and  extrapolating at higher  energy.
In the figures we  show  two calculations  based
on   equation (\ref{eq:wtot})  using (as  in the previous  figure)
the constant  values $w=3$ and $r_0 = 0.19$~fm, and the two  parametrizations
of  $\sigma_{\rm eik}(s)$
(the results for  the model  are only plotted for  $\sqrt{s} > 60$~GeV).

Figure~\ref{fig:stot} also  shows  the parametrizations
for  $\sigma_{\rm tot}^{pp} (s)$
and   $\sigma_{\rm tot}^{\overline{p}p} (s)$ suggested
in the PDG \cite{cudell,pdg}.   The PDG  estimate of the
extrapolation of the total $pp$ cross sections falls  in between 
our two   estimates,  that mark a  range of   uncertainty
in our prediction.

At the LHC  energy  ($\sqrt{s} = 14$~TeV) the PDG  prediction 
is  $\sigma_{\rm tot} = 112.2$~mbarn
while  our  two calculations  give
$\sigma_{\rm tot} = 98.1$ and 120.8~mbarn
[at $\sqrt{s} = 10$~TeV  the PDG  predictions 
is  $\sigma_{\rm tot} = 105.7$~mbarn,
 our  calculations  give
94.2  and 112.7~mbarn].
The extrapolation  to $\sqrt{s} = 4.33\times 10^{5}$~GeV  (that corresponds
to a proton cosmic  ray particle with energy $E_{\rm lab} = 10^{20}$~eV
is  $\sigma_{\rm tot} = 194$~mbarn  for the PDG extrapolations,
and 146 and 229 ~mbarn for our  calculations.

Recently two  groups  \cite{Ryskin:2007qx,Gotsman:2008tr}
have discussed  predictions of  the total $pp$ cross section  
at high energy  in the  framework of  models  that
include  a  treatment of diffraction  with a multi--channel eikonal
and  multi--pomeron  interactions.
Both  groups  arrive to a similar conclusion,  namely that 
the inclusion of diffraction  reduces  the estimate of the total cross section
at LHC  energy.
For  the two  groups  the estimate of  
$\sigma_{\rm tot}$ at $\sqrt{s} = 14$~TeV is of order 90~mbarn, 
approximately $20$\%  smaller than  the PDG.
Their estimates of the total cross section  grows very slowly with energy
reaching   $\sigma_{\rm tot} \simeq 108$~mbarn (for \cite{Gotsman:2008tr})
and $\sigma_{\rm tot} \simeq 98$~mbarn (for \cite{Ryskin:2007qx})
at $\sqrt{s}  = 10^{5}$~GeV  (that corresponds to $E_{\rm lab} = 5.33 \times 10^{18}$~eV).

The main point that we  want to make here, is that it is
certainly the case that  given a model for  
$\langle n(b,s) \rangle$  (that in  the simple
eikonal model  is simply  equal to twice  the eikonal  function
$\chi (b,s)$),  the inclusion of  diffraction  reduces  the cross section.
However, the estimate  of  the total  cross section and of  
its  dependence on  energy also  involves  the  calculation  of the function
$\langle n(b,s) \rangle$. 

The conclusion that the cross section at LHC  
is of order 90~mbarn obtained  by the authors  in 
 \cite{Ryskin:2007qx,Gotsman:2008tr} 
should not be considered as a  consequence  of the
inclusion of  diffraction  in the theoretical  framework, but rather as
the consequence of  the entire set of theoretical assumptions of their  models.

\begin{figure}[hbt]
\begin{center}
 \includegraphics[angle=90,width=15cm]{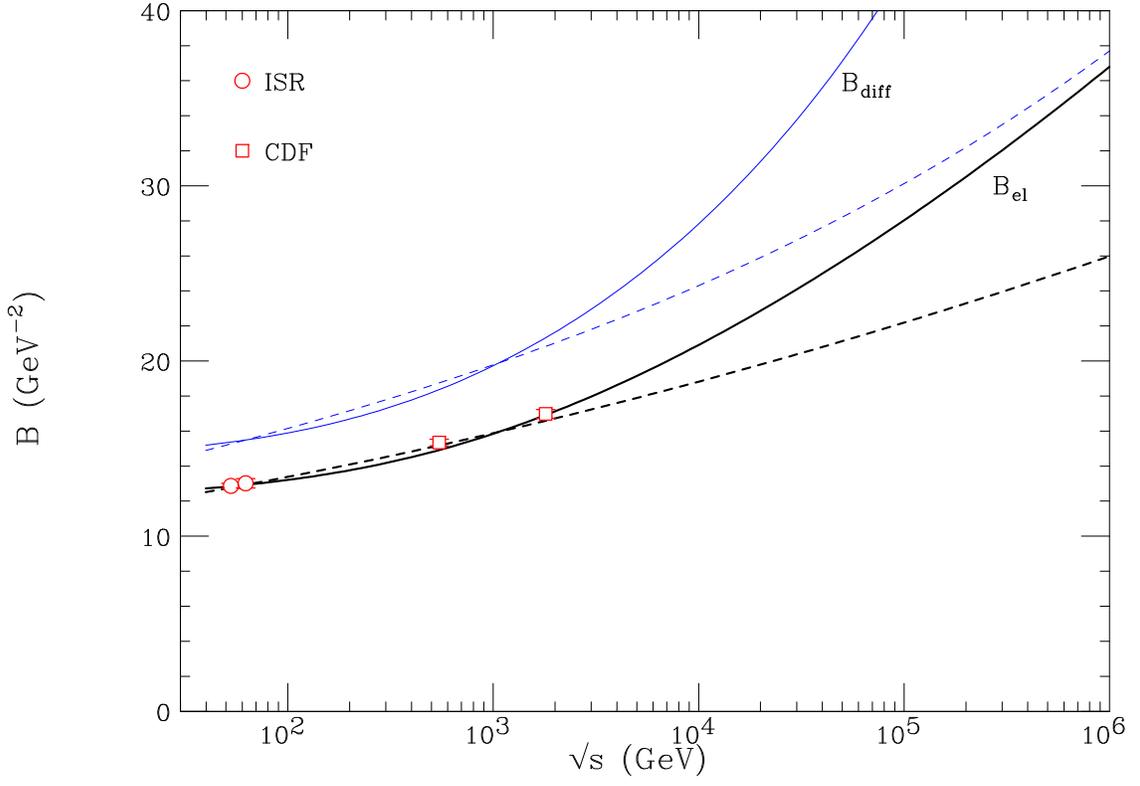}
\end{center}
   \caption{\small  
Slope at $t=0$  of  the differential  cross
sections  $d\sigma_{\rm el}/dt$  and 
 $d\sigma_{\rm diff}/dt$  for 
  elastic  and inelastic  diffractive  events. 
The  predictions  are calculated
with equations 
(\protect\ref{eq:diffr1})
and (\protect\ref{eq:diffr2}), using 
the functional form  (\protect\ref{eq:p_model})  for
$p(\alpha)$  with $w=3$,  and the parametrization (\protect\ref{eq:pareik}) for
$\langle n(b,s)\rangle$  with  $r_0 = 0.19$~fm.
The solid  and dashed  curves use the two parametrizations
of $\sigma_{\rm eik}(s)$ shown in  fig.~\protect\ref{fig:fit_seik}.
 \label{fig:bslope}  }
\end{figure}

Figure~\ref{fig:bslope}  shows, 
plotted  as  a function of  $\sqrt{s}$,
the   predictions of our  model  for the 
slopes  $B_{\rm el} (s)$ and $B_{\rm diff}(s)$  
of the differential
cross sections  $d\sigma_{\rm el}/dt$  and 
 $d\sigma_{\rm diff}/dt$.  
For each slope,  the  figure shows  two  curves that  differ   for the use
of  the two different   parametrizations  of $\sigma_{\rm eik} (s)$
that are shown in  fig.~\ref{fig:fit_seik},  and  have already been used
in fig.~\ref{fig:diffraction}  and~\ref{fig:stot}.
Note  how  $B_{\rm  diff}(s)$ is  always  larger  than $B_{\rm el}(s)$.

\section{Summary and Outlook}
\label{sec:concl}
In this work we have discussed the problem of multiple parton
interactions in hadron  collisions.
If one takes into  consideration  only  the parton 
scatterings that have sufficiently large momentum transfer,
it becomes  possible to 
detect the  final  state  partons  as  high  $p_\perp$  jets,  and  determine
event  by  event the  number of hard parton interactions  that  are present.
It  becomes  therefore  possible to study  experimentally  the 
multiplicity  distribution of   parton interactions
above for  example $p_\perp^{\rm min}$.
At sufficiently high c.m. energy the probability
of having more than one high $p_\perp$ parton scattering
in a single collision can be appreciable,  
and the hard scattering multiplicity  distribution 
becomes non trivial.

The calculation for  the inclusive distribution of   high  $p_\perp$ 
parton scattering  is  a  textbook application  of perturbative QCD
and  can be performed  from   a knowledge of the  standard PDF's.
A theoretical prediction of the  multiplicity  distribution of 
high $p_\perp$ interactions in high energy hadron collisions
is however a highly non trivial problem
that requires the introduction of additional  theoretical
concepts.

The  standard PDF's   give  the 
inclusive probability    density for  finding one single  parton  with   fractional
longitudinal momentum  $x$. 
This probability density is obtained 
integrating over the parton transverse momentum,
and  integrating over all possible momenta
of  the other partons in the hadron.
If one  wants  to compute the  probability to  have
exactly $n$  hard interactions  in one  collision
the information contained  in the PDF's is clearly insufficient.
One  needs  to  know:
\begin{enumerate}
\item  the probability to find
a parton of a given $x$  at different  impact parameters with respect
to the  hadron  center of  mass; 
\item  the  {\em correlated} probabilities for finding
different  partons   at
 ($x_1,\vec{b}_1$), 
 ($x_2,\vec{b}_2$), 
 ($x_3,\vec{b}_3$) , $\ldots$. 
\end{enumerate}
The  first problem  should be addressed  introducing
impact parameter  dependent  PDF's, 
$F_j^h(x,b, Q^2)$  that  give  the probability of  finding
 the parton  of  type $j$  with fractional longitudinal  momentum  $x$ and
impact parameter $\vec{b}$  probing  hadron  $h$ at the  scale $Q^2$.
This problem   has  not  yet  a  well  determined  solution,
and  all studies  of this problem have made  the simplification
to assume  that the  dependences  on $x$ and $b$ of the impact parameter
PDF's   factorize, that is:
$F_j^h(x,b, Q^2)  = f_j (x,Q^2) ~\hat{\rho}(b)$,
and  estimated  $\hat{\rho} (b)$ with simple  phenomenological considerations.
Studies of the Generalized PDF's
\cite{Burkardt:2002hr,Diehl:2003ny,Belitsky:2005qn}
 should  soon be  able to  shed light on this  question. 

The problem of obtaining correlated PDF's that give
the  probability to  find simultaneously several partons
in different  elements  of phase  space is  clearly 
much more difficult and complex.
In this work we have suggested to parametrize the 
effects of our lack of knowledge about the correlated 
PDF's introducing 
the ``effective  configuration  probability distribution'',  that is 
one  function $p(\alpha)$   the  real, positive variable $\alpha$.
Each one of the configurations of partons  in 
the pair of  colliding hadron has associated  the  real number 
$\alpha$. The  physical  meaning  of $\alpha$ is that the
expected number of  parton interactions  that corresponds
to the parton configuration  ${\mathbb{C}}$ is
$n({\mathbb{C}}) = \langle n\rangle \; \alpha$,  where $\langle n\rangle$ is
the  average  over all configurations. 
The  first two moments of the function $p(\alpha)$  are unity  (because of the
normalization of a probability density and  to 
reproduce the correct $\langle n \rangle$);  increasing  the 2nd moment of
the $p(\alpha)$  distribution 
the width of the  multiplicity  of parton interactions  grows.

If one considers not only a  subset of  detectable  (high $p_\perp$) 
parton  interactions, but {\em all}  of them,  it  becomes
possible to relate these  elementary  
interactions  with the total  and elastic cross sections.
This  general idea  has  been implemented  in  many  works  using
an eikonal  formalism.  A  crucial  ingredient  of  these models
is  the quantity  $\langle n(b,s)\rangle$  that  gives   the 
average  number of elementary interactions  for a hadron collision
at impact parameter $b$ and c.m.  energy $\sqrt{s}$.
In the simple  eikonal model this  quantity is  related  to the elastic
scattering   profile  function by the relation
$\Gamma_{\rm el} (b,s) = 1 - \exp[-\langle n(b,s)\rangle/2]$.
The corresponding  inelastic  cross section is   then:
$$
\sigma_{\rm inel} (s) = \int d^2 b ~\left \{1 - \exp[-\langle n(b,s)\rangle ]
\right  \}
$$
The  physical  interpretation is   that    an inelastic interaction
corresponds  to   absorption and to at least one elementary interaction,
assuming  Poisson  fluctuations in their multiplcity.
The same   considerations  that we have  outlined    for  hard interactions  however
apply, and it is  natural to expect that  fluctuations 
in the number of  elementary interactions $n$  are in fact much  broader
than poissonian  because of  fluctuations in the  ``configurations''
of  the colliding  hadrons.  
This effect  can again be  parametrized with a function $p(\alpha)$.
For example the inelastic cross section can  be  rewritten as:
$$
\sigma_{\rm inel} (s) = \int d^2 b ~
\int_0^\infty d\alpha~p(\alpha) 
~\left \{ 1 - \exp[-\langle n(b,s)\rangle \; \alpha  ] \right \}
$$

One can  see that the parameter $\alpha$ 
controls the ``transparency''  of a hadron collision.
Different ``configurations''  of  the colliding  hadrons  have
transparencies  that are  related to $\alpha$.
Good and  Walker \cite{Good:1960ba}  have proposed that inelastic  diffraction
originates  from the different  absorption 
of  the  different  components  of the colliding hadrons.
Therefore   our formalism can be applied 
to the calculation of  the inelastic  diffractive cross section,  and in fact
unavoidably implies  the presence  of inelastic  diffractive  processes.

In other words,  the function $p(\alpha)$ allows to
relate the   quantity $\langle n(b,s) \rangle$   to the total and
elastic  cross sections, and at the same  time  fixes
the value of   the diffractive cross section.
Viceversa,  from the data  on the total and  elastic  cross
section,  together with the data on inelastic  diffraction it is possible
to  extract information on $\langle n(b,s) \rangle$ and on the properties
of $p(\alpha)$.

We  have performed an analysis of the data  on
$pp$ and  $\overline{p}p$  collisions  obtained  at high energy
colliders, and  obtained  information 
on $\langle  n(b,s)\rangle$ and  $p(\alpha)$.
For the study of the properties  of  $p(\alpha)$ we have used as
a  first approximation a simple  analytic  form  that
depends on a single parameter.

To describe  the measured    diffractive cross section one
is  forced to  have a fuction $p(\alpha)$ with  a  large
variance. This  in  turn  has  very important  consequence
on the parameters  that  describe  $\langle n(b,s) \rangle$.
It is  remarkable  that we  find  that 
(within   significant uncertainties)  the function 
$p(\alpha)$ is    independent from energy; moreover
parametrizing  $\langle n(b,s) \rangle$ in the form:
$\langle n(b,s) \rangle = \sigma_{\rm eik} (s) \; A(b,s)$
as  the product of  an  eikonal  cross section times
a  geometrical overlap  function,  we find that
the   geometrical factor  can  be taken  as energy independent,
in   contrast  with results  obtained in  the simple eikonal model
that  neglects  fluctuations.
The eikonal   cross  section $\sigma_{\rm eik}(s)$ is much larger
and grows much faster with energy than in 
the  simple  eikonal  model.   Such a  rapid  growth can   however 
be readily explainable  assuming  that it is controled
by the  increase of $\sigma_{\rm jet}(p_\perp^{\rm min}, s)$
with $s$ assuming  a constant  $p_\perp^{\rm min}$ and
negligible  screening  effects.

We note  that, at least in first approximation,
the function $p(\alpha)$ that  we have  extracted 
from the  study of  the total, elastic  and  diffraction
cross sections, is also applicable to the study of the  multiplicity
distribution of high $p_\perp$ jets.  
One can therefore make the
prediction that the distribution  
of the number of hard interaction per event  will be
broad, with a non negligible number of  events  containing
several interactions.

The  prediction of the total cross section  at LHC 
depends on the  energy dependence  of $\sigma_{\rm eik} (s)$.
This  is  a problem  we have not discussed in  detail here. 
It seems however natural  to  expect a  result around
110~mbarn  with however  a  significant  uncertainty.
In the model  we  are discussing however
the  eikonal  cross section $\sigma_{\rm eik}(s)$ is large
and  since   $\sigma_{\rm eik}(s)/\sigma_{\rm  inel}(s)$  is  equal to the number
of elementary interactions in a  collision one is  lead to expect 
a  large charged particle  multiplicity  and a soft inclusive spectrum
in the final  state.  These consideration are  also  relevant for the study of
ultra  high energy cosmic ray showers.

\vspace{0.3 cm}
\noindent{\bf Acknowlegments}  We  are  grateful to  Gianni Salm\'e
for  discussions  about the GPDF's. One of us  (PL)  would
like to  thank Ralph Engel  for the  introduction  to  the model
of appendix~\ref{sec:4chan}, and 
Tom  Gaisser,  Todor  Stanev  and  Eun--Joo Ahn
for many discussions  about  hadron interactions.

\appendix

\section{Electromagnetic form  factors    and overlap function $A(b)$}
\label{sec:overlap}
In  \cite{Durand:1988cr}   Durand and Pi  estimated
the  overlap function $A(b)$   for  $pp$ collisions 
from the electromagnetic  form  factor  of the proton. 
The  simple  physical  idea  behind their    derivation  is that
the overlap function   is   the  energy independent 
geometric  overlap  of the   hadronic  matter  distributions  in the
colliding particles.
More  explicitely,  one defines
the spatial  distribution  of  matter in the hadron $h$ as
$\rho_h(r)$,  with the  normalization condition:
\begin{equation}
\int d^3 r~ \rho_h (r) = 1
\label{eq:rhonorm}
\end{equation}
The density   in  the transverse plane is then  be obtained 
with a  simple integration:
\begin{equation}
\hat{\rho}_{h} (b) = \int_{-\infty}^{+\infty} dz ~\rho_{h} \left (
\sqrt{ b^2 + z^2}\right )
\end{equation}
The  overlap function  in the collision 
between  hadrons $h_1$ and  $h_2$  is then obtained as:
\begin{equation}
A(b) = \int d^2b_1 ~d^2b_2 ~\hat{\rho}_{h_1}(\vec{b}_1) 
~\hat{\rho}_{h_2} (\vec{b}_2)~\delta [\vec{b} - \vec{b}_1 + \vec{b}_2]
\end{equation}
The   normalization  condition
\begin{equation}
\int d^2b ~A (b) = 1
\end{equation}
follows  automatically   from 
the  normalization of  $\rho_{h_1}(r)$ and  $\rho_{h_1}(r)$ given by (\ref{eq:rhonorm})

To  estimate the density of $\rho_h(r)$ of  hadron $h$,
 Durand and Pi make the assumption that 
it is  simply  the Fourier 
transform  of its  electromagnetic  form factor.
For the  proton one   has:
\begin{equation}
F_p (q^2) = \frac{1}{(1 + R_p^2 \, q^2)^2}
\label{eq:formp}
\end{equation}
with  $R_p = 0.234$~fm  (or $R_p^{-2} = 0.71$~GeV$^2$)  and correspondingly 
$\rho_p (r) \propto  e^{-r/R_p}$.  The  geometric  convolution for  proton--proton
collisions is:
\begin{equation}
A_{pp} (b) = \frac{b^3}{96 \, \pi \, R_p^5}  \; K_3 \left ( \frac{ b}{R_p} \right )
\label{eq:App0}
\end{equation}
where $K_3(x)$ is the  modified  Bessel  function of the third kind.

\section{Four channels  Model}
\label{sec:4chan}
It is  instructive  to discuss a ``minimum''  model that
implements  the Good--Walker  
ansatz \cite{Good:1960ba}   for inelastic  diffraction
in the collision  between two hadrons, 
where all  calculations
can be performed   explicitely. 
The  minimum  model  has  4 channels,
to describe the 4  possible types   of scattering
(target, projectile  and double diffraction  together with elastic scattering).
Without loss of  generality
we can consider the scattering  $\pi p$  
(here  ``$\pi$'' and ``$p$''  are  labels  to represent
arbitrary hadrons).  Each of the two  colliding  hadrons  can
undergo    inelastic  diffraction  with  a   transition to
an additional  state. We will label the excited states
for  the projectile  and target particles  as ``$\rho$'' and
``$\Delta$''; one  therefore has  to consider the  transitions:
$\pi \to \pi^* \equiv \rho$ and $p \to p^* \equiv \Delta$. 
In the 4--channel  model one    has to study the  4--dimensional
vector space  spanned  by  the  orthonormal  basis
of the 4~physical states $|\varphi_m\rangle$:
\begin{equation}
\{|\varphi_m\rangle\}_{(m=1,4)}  = 
\{
|\pi p\rangle , 
|\pi \Delta\rangle, 
|\rho p\rangle , 
\rho \Delta\rangle
\}
\end{equation}
One can (in principle)  study   the   $4 \times 4$   transitions  
$\langle \varphi_f | \mathbf{S} | \varphi_i\rangle$.
In practice of course one  is limited to the study  of the
transitions $|\pi p\rangle \to |\varphi_m\rangle$ 
(that  correspond to  the processes of 
elastic scattering, target, projectile  and double
diffraction). 
The scattering  amplitude  is a $4 \times  4$ matrix: 
\begin{equation}
\hat{F} (\vec{q}, s) =  i ~ 
\int \frac{d^2b}{2 \, \pi} 
~e^{i \, \vec{q}\cdot \vec{b}}
~\hat{\Gamma} (b,s) 
\end{equation}
The differential cross section for the  transition $i \to f$  is:
\begin{equation}
\frac{d\sigma_{i \to f}}{dt} (t, s)= \pi ~ \left|\left [\hat{F}(\vec{q},s) 
\right]_{fi}\right |^2
\end{equation}
Integrating  over all  $t$  values one obtains the 
transition   cross sections
\begin{equation}
\sigma_{i \to f}(s)= 
\int d^2b ~\left |\left [\hat{\Gamma} (b,s) \right ]_{fi} \right |^2 
\end{equation}
The  profile  matrix $\hat{\Gamma}$  can be written
in terms of   the  eikonal  matrix $\hat{\chi}(b,s)$:
\begin{equation}
\hat{\Gamma} (b,s) = 1 - \exp \left [- \hat{\chi} (b,s) \right ]~.
\end{equation}
Using the Good and Walker ansatz, the  eikonal   matrix  $\hat{\chi}(b,s)$
takes the form:
\begin{equation}
\hat{\chi}(b,s) = \frac{\langle n (b,s) \rangle}{2} ~ \hat{M}
\end{equation}
where  $\langle n(b,s) \rangle$   has the usual  meaning
of the average  number of parton interactions  in a $\pi p$  collision,
and   we  have introduced
the $4 \times  4$ matrix  $\hat{M}$. 
This matrix   must be  real  and  have  4  real  and
positive  eigenvalues $\alpha_j$,   
moreover  one must have   $\hat{M}_{11} = 1$.

This is a  consequence of the   fact that 
one  can  define
(as  in the previous  section)  the  states $|\psi_j\rangle$   as the  
eigenstates of  the $\hat{M}$ matrix.  These  states
undergo  only absorption  or  elastic scattering,
and each  has a  ``transparency'' 
$P_0 = e^{-\langle n(b,s) \rangle \, \alpha_j}$, therefore
$\langle n(b,s) \rangle \; \alpha_j$ can be interpreted as
the average number of interactions  for the
state $|\psi_j\rangle$   and  therefore this quantity
(and $\alpha_j$) must  be positive.  
Moreover, to  have the correct average  multiplicity of  elementary 
interactions  for the  intiali state $|\varphi_1\rangle = |\pi p\rangle$
one must have:
\begin{equation}
\sum_j  \left | \langle \psi_j | \phi_1 \rangle \right |^2 \; \alpha_j = 1
\end{equation}
that implies   $\hat{M}_{1,1} = 1$.

The  matrix $\hat{M}$ can be  constructed  explicitely
making the additional  hypothesis that  the (4--dimensional) space of the
physical  states is  the  direct product 
of  two  (2--dimensional)  spaces    
for the  beam and target particle, and moreover
that  one  has   time  reversal   symmetry,  and  the amplitude for
the transitions  
 $\pi \to  \rho$   ($p \to \Delta)$
and 
 $\rho \to  \pi$  ($\Delta \to p$)   are equal.
With this assumptions  the most  general  form for the 
matrix  $\hat{M}$ is: 
 
\begin{eqnarray}
\hat{M} & =  & 
\begin{pmatrix} 
1 & \beta_\pi  \\
\beta_\pi & 1 - 2 \epsilon_\pi
\end{pmatrix} 
~\otimes ~
\begin{pmatrix} 
1 & \beta_p  \\
\beta_p & 1 - 2 \epsilon_p
\end{pmatrix} 
\nonumber \\
& ~ & \nonumber \\
& = & 
\begin{pmatrix} 
1 & \beta_\pi & \beta_p & \beta_\pi \, \beta_p \\
\beta_\pi & 1 \,-\, 2 \epsilon_\pi & \beta_\pi \, \beta_p & \beta_p \, (1 \,-\, 2 \epsilon_\pi) \\
\beta_p & \beta_\pi \, \beta_p & 1 \,-\, 2 \epsilon_p & \beta_\pi \, (1 \,-\, 2 \epsilon_p) \\
\beta_\pi \, \beta_p ~ & \beta_p \, (1 \,-\, 2 \epsilon_\pi) ~ & \beta_\pi \, (1 \,-\,
2 \epsilon_p) ~& (1 \,-\, 2 \epsilon_\pi) \, (1 \,-\, 2 \epsilon_p)
\end{pmatrix}
\end{eqnarray}

The  eigenvalues and eigenvectors  of the  matrix $\hat{M}$ are easily 
calculable, noting   that each  of the $2 \times 2$ matrices
of  form
\begin{equation*}
\begin{pmatrix} 
1 & \beta  \\
\beta & 1 - 2 \, \epsilon
\end{pmatrix} 
\end{equation*}  
has   eigenvalues:
\begin{equation}
\lambda_{1,2} = 1 \pm \gamma - \epsilon
\end{equation}
where  
\begin{equation}
 \gamma = \sqrt{\beta^2 + \epsilon^2}~.
\end{equation}
and the  corresponding eigenvectors are:
%\begin{equation}
%\vec{v}_{1,2} = \left \{ 
%\pm \sqrt{ 
%\frac{ \gamma \pm \epsilon } {2 \; \gamma}},
% \sqrt{\frac{\gamma \mp \epsilon } {2 \; \gamma}} \right \}
%\end{equation}
\begin{equation}
\vec{v}_{1,2} = \frac{1}{\sqrt{2}} \;  \left \{ 
\pm \sqrt{1 \pm r}, ~ 
 \sqrt{1 \mp r} \right \}
\end{equation}
with $r = \epsilon/\gamma = \epsilon/\sqrt{\beta^2 + \epsilon^2}$.

The eigenvalues of the $4 \times 4$ matrix  $\hat{M}$ are then:
\begin{equation}
\alpha_{j} = (1 \pm \gamma_\pi - \epsilon_\pi) \; (1 \pm \gamma_p - \epsilon_p)
\end{equation}
(with $j\in \{1,2,3,4\}$).
The condition that  the  eigenvalues are non--negative gives:
\begin{equation}
\epsilon_{\pi,p} \le 1/2; ~~~~~ \beta^2_{\pi, p} 
\le 1 - 2 \epsilon_{\pi,p}
\end{equation}
The rotation matrix  $C_{m j}$   that  connects
the  scattering  eigenstates  $|\psi_j\rangle$ to the physical  states $|\varphi_m\rangle$ 
is:
\begin{equation}
C_{m j} = 
\frac{1}{2}  ~
\begin{pmatrix} 
\sqrt{1 + r_\pi} \, \sqrt{1+ r_p}~~~ &
\sqrt{1 - r_\pi} \, \sqrt{1+ r_p}~~~ &
\sqrt{1 + r_\pi} \, \sqrt{1- r_p}~~~ &
\sqrt{1 - r_\pi} \, \sqrt{1- r_p}~~~  \\

-\sqrt{1 - r_\pi} \, \sqrt{1+ r_p}~~~ &
\sqrt{1 + r_\pi} \, \sqrt{1+ r_p}~~~ &
-\sqrt{1 - r_\pi} \, \sqrt{1- r_p}~~~ &
\sqrt{1 + r_\pi} \, \sqrt{1- r_p}~~~  \\

-\sqrt{1 + r_\pi} \, \sqrt{1- r_p}~~~ &
-\sqrt{1 - r_\pi} \, \sqrt{1- r_p}~~~ &
\sqrt{1 + r_\pi} \, \sqrt{1+ r_p}~~~ &
\sqrt{1 - r_\pi} \, \sqrt{1+ r_p}~~~  \\

\sqrt{1 - r_\pi} \, \sqrt{1- r_p} &
-\sqrt{1 + r_\pi} \, \sqrt{1- r_p} &
-\sqrt{1 - r_\pi} \, \sqrt{1+ r_p} &
\sqrt{1 + r_\pi} \, \sqrt{1+ r_p}  \\
\end{pmatrix} 
\end{equation}

The profile  functions for the    different  scattering processes can
now  be given  explicitly as:
\begin{equation}
\Gamma_{m_f \, m_i} = \sum_j C_{m_f  \, j} \; C_{m_i  \, j} ~\left [1 - 
\exp \left ( - \frac{\langle n (b,s) \rangle }{2} \, \alpha_j \right ) \right ]
\end{equation}
The  model  outlined above     requires  an estimate of 
the  function $\langle n (b,s) \rangle$ 
that can be interpreted as the average  number  of ``elementary''  interactions
for a  hadron crossing  at impact parameter $b$ and  c.m. energy $\sqrt{s}$.
In  the general  case
of  the collision of  two different hadrons
(such as in $\pi^\pm p$ scattering) the model  has   4 additional 
parameters  ($\beta_\pi, \epsilon_\pi, \beta_p, \epsilon_p$)
 that describe the matrix structure of the eikonal function.
Obviously  for $pp$ scattering   the model  has  
only two parameters     ($\beta_p$, $\epsilon_p$).

With the labeling of the physical  states that we have been using
(namely: 
$|\varphi_1\rangle = |\pi p\rangle$,
$|\varphi_2\rangle = |\pi \Delta\rangle$,
$|\varphi_3\rangle = |\rho p\rangle$ and
$|\varphi_4\rangle = |\rho \Delta\rangle$)
the integration over all  impact  parameters $b$ of 
$|\Gamma_{11}(b)|^2$  yields the  elastic cross section,
the integral  of  $|\Gamma_{21}(b)|^2$  
 ($|\Gamma_{31}(b)|^2$)  gives  the  target  (projectile)  single
diffraction cross section, and finally  the integral 
of $|\Gamma_{41}(b)|^2$  gives  the double  diffractive cross section.  

To  connect  this  analysis  to the discussion performed  in the main  text,
we  note  that we  can    define the  4 quantities $p_j$ 
that are the  probabilities 
$|\langle \varphi_1 | \psi_j\rangle |^2$ 
to find the   initial state   $|\varphi_1 \rangle \equiv |\pi p\rangle$   
in  the scattering  eigenstates $|\psi_j\rangle$.
The  $p_j$  are given  by:
\begin{equation}
p_j = \left | \langle \psi_j|\pi p\rangle \right |^2  =
\frac{(\gamma_\pi \pm \epsilon_\pi) \;(\gamma_p \pm \epsilon_p) }
{4 \, \gamma_\pi \; \gamma_p} 
= \frac{1}{4} ~ (1 \pm r_\pi) \; (1 \pm r_p)
\end{equation}
It is  straightforward to verify that:
\begin{equation}
\sum_j  p_j = \sum_j p_j \; \alpha_j = 1 
\label{eq:pdiscr}
\end{equation}
One   can  now  define the  function $p(\alpha)$:
\begin{equation}
p(\alpha) = \sum_j p_j \; \delta [\alpha - \alpha_j]
\end{equation}
This function, as a consequence of equations (\ref{eq:pdiscr})
satisfies the conditions:
$$
\int_0^\infty d\alpha \; p(\alpha) = 1, ~~~~~~~
\int_0^\infty d\alpha \; \alpha~ p(\alpha) = 1.
$$

It is now  straightforward to see  that one can recast the expressions
for the  total, elastic, absorption and diffractive  
(that is the sum of the target, projectile and double  diffraction) cross section
as integrals over $\alpha$  identical to  the expressions  
(\ref{eq:sig_abs}--\ref{eq:sig_tot}) in section~\ref{sec:partonic}.

It can be interesting to note  that the  2nd moment  of the $p(\alpha)$  distribution 
is  given by:
\begin{equation}
\int d\alpha ~\alpha^2 ~p(\alpha) = \sum_j p_j ~\alpha_j^2  = (1 + \beta_p)\, (1+\beta_\pi)
\end{equation}

\clearpage

\clearpage

\end{document}